\documentclass{aa}  

\usepackage{graphicx}
\usepackage{subcaption}
\usepackage{booktabs}
\usepackage{amsmath}
\usepackage{lipsum}
\usepackage{comment}
\usepackage{adjustbox}
\usepackage{multirow}
\usepackage{txfonts}
\usepackage{float}
\usepackage{natbib}
\usepackage[hypertexnames=false]{hyperref}
\usepackage{lscape}
\usepackage{placeins}
\usepackage{textgreek}
\usepackage{xspace}
\usepackage{xargs}
\usepackage[usenames,dvipsnames]{xcolor}
\usepackage{enumitem}
\usepackage{pifont}

\newcommand{\comm}[1]{{\color{teal}[#1]}}

\newcommandx{\mgii}[1][1=]{\textup{Mg\,\textsc{ii}}{#1}\xspace}
\newcommandx{\niii}[1][1=]{\textup{N\,\textsc{iii}]}{#1}\xspace}
\newcommandx{\niv}[1][1=]{\textup{N\,\textsc{iv}]}{#1}\xspace}
\newcommandx{\nv}[1][1=]{\textup{N\,\textsc{v}}{#1}\xspace}
\newcommandx{\cii}[1][1=]{\textup{[C\,\textsc{ii}]}{#1}\xspace}
\newcommandx{\ciii}[1][1=]{\textup{C\,\textsc{iii}]}{#1}\xspace}
\newcommandx{\civ}[1][1=]{\textup{C\,\textsc{iv}}{#1}\xspace}
\newcommandx{\oiii}[1][1=]{\textup{[O\,\textsc{iii}]}{#1}\xspace}
\newcommandx{\oiv}[1][1=]{\textup{O\,\textsc{iv}}{#1}\xspace}
\newcommandx{\siiv}[1][1=]{\textup{Si\,\textsc{iv}}{#1}\xspace}
\newcommandx{\feii}[1][1=]{\textup{Fe\,\textsc{ii}}{#1}\xspace}
\newcommandx{\lya}[1][1=]{\textup{Ly\,\textsc{$\alpha$}}{#1}\xspace}

\newcommand{\vel}{\ensuremath{\mathrm{v}_{98}}}
\newcommand{\velred}{\ensuremath{\mathrm{v}_{\rm 98,red}}}

\newcommand{\vminBAL}{\ensuremath{\mathrm{v}_{\rm min,BAL}}}
\newcommand{\vmaxBAL}{\ensuremath{\mathrm{v}_{\rm max,BAL}}}
\newcommand{\deltav}{$\rm \Delta v$}
\newcommand{\LEdd}{$\rm \lambda _{Edd}$}

\begin{document}

   \title{Multi-tracer constraints on quasar outflows}

   \subtitle{Acceleration from BLR to galactic scales}

%%%%%%%%%%%%%%%%%%%%%%%%%%%%%%%%%%%%%%%%
% Please separate each author with the \and command
%
% Please do not include ORCIDs next to author names.
% Only ORCIDs authenticated by individual authors in EDPS
% editorial system will be taken into account.
% ORCIDs included here will be removed.
%%%%%%%%%%%%%%%%%%%%%%%%%%%%%%%%%%%%%%%%

   \author{M. Brazzini\inst{1,2,3}
        \and M. Bischetti\inst{4,2,3}
        \and S. Hagen\inst{3,5,2}
        \and V. D'Odorico\inst{2,3}
        \and N. Arav\inst{6}
        \and G. Becker\inst{7}
        \and G. Cupani\inst{2,3}
        \and F. Fiore\inst{2,3}
        \and M. Ghosh\inst{6}
        \and E. Piconcelli\inst{8}
        \and F. Salvestrini\inst{2,3}
        \and M. Sharma\inst{6}
        \and A. Travascio\inst{2}
        \and A. Tortosa\inst{9}
        \and G. Vietri\inst{10}
        \and L. Zappacosta\inst{8}
        \and C. Feruglio\inst{2,3}
        }

   \institute{
   Department of Physics, Astronomy Section, University of Trieste, Via G.B. Tiepolo, 11, I-34143 Trieste, Italy, \email{matilde.brazzini@phd.units.it}  %1
   \and INAF - Osservatorio Astronomico di Trieste, Via G. B. Tiepolo 11, I-34143 Trieste, Italy, \email{matilde.brazzini@inaf.it}  %2
   \and IFPU - Institute for Fundamental Physics of the Universe, Via Beirut 2, I-34014 Trieste, Italy %3
   \and Dipartimento di Fisica ”Enrico Fermi”, Università di Pisa, Largo Bruno Pontecorvo 3, Pisa, I-56127, Italy %4
   \and SISSA – International School for Advanced Studies, Via Bonomea 265, 34136 Trieste, Italy %5
   \and Department of Physics, Virginia Tech, Blacksburg, VA 24061, USA %6
   \and Department of Physics \& Astronomy, University of California, Riverside, CA 92512, USA %7
   \and INAF - Osservatorio Astronomico di Roma, Via Frascati 33, 00040 Monte Porzio Catone, Rome, Italy %8
   \and INAF - Osservatorio di Astrofisica e Scienza dello Spazio di Bologna, via Piero Gobetti, 93/3, I-40129 Bologna, Italy %9
   \and INAF – Istituto di Astrofisica Spaziale e Fisica Cosmica Milano, Via Alfonso Corti 12, Milano, 20133, Italy %10
   }

   \date{Received September 30, 20XX}

% \abstract{}{}{}{}{}
% 5 {} token are mandatory
 
  \abstract
  % context heading (optional)
  % {} leave it empty if necessary  
   {Black hole-driven outflows are a fundamental component of the Active Galactic Nuclei (AGN) paradigm, and may play a major role in regulating the evolution of both supermassive black holes and their host galaxies. Yet, their physical properties remain challenging to constrain, due to their multi-phase and multi-scale nature which demands a combined approach of different complementary observational tracers. }
  % aims heading (mandatory)
   {In this work, we investigate AGN-driven ionized outflows from nuclear to galaxy scale within individual quasars, with the aim of establishing whether and how these phenomena are physically connected.} 
  % methods heading (mandatory)
   {We assemble a luminosity-matched sample of $148$ quasars with $\rm L_{bol}>10^{47}$~erg s$^{-1}$ spanning the redshift range $z \sim 2-6.5$. We perform a homogeneous analysis of \civ emission and broad absorption line (BAL) profiles to derive outflow properties on different spatial scales. }
  % results heading (mandatory)
   {We find statistically significant correlations between emission- and absorption-based outflow diagnostics, including the minimum and maximum BAL velocities, \vminBAL\ and \vmaxBAL, and the \vel\ of \civ emission lines. These correlations are tighter in high redshift quasars. We also find evidence for an evolution of outflow properties across cosmic time, with high-redshift quasars exhibiting systematically larger \civ line shifts and broader profiles, and faster BAL outflows. According to our outflow toy model, these correlations suggest a scenario in which a clumpy BAL wind is composed of expelled and accelerated broad-line region gas clouds.}
   {}

   \keywords{Galaxies: active -- (Galaxies:) quasars:general -- (Galaxies:) quasars: supermassive black holes
               }

   \maketitle

\nolinenumbers

%%%%%%%%%%%%%%%%%%%%%%%%%%%%%%%%%%%%%%%%%%%%%%%%%%%%%%%%%%%%%%
\section{Introduction}
Black hole-driven outflows are a key component of the active galactic nuclei (AGN) paradigm, representing the primary feedback process shaping the evolution of both supermassive black holes (SMBHs) and their host galaxies across cosmic time \citep[e.g.,][]{Alexander+Hickox2012}.
In particular, quasars, the most luminous class of AGN, are expected to inject large amounts of energy and momentum into their surroundings through powerful winds capable of substantially regulating the successive evolution of the environment they are embedded in \citep{Fabian2012, Kormendy&Ho2013, King+2015}.
Despite their importance, a comprehensive characterization of these winds remains challenging due to their intrinsic multi-phase and multi-scale nature \citep{Harrison2024}.
Although their broad phenomenology is now well established, several fundamental questions remain open, including how the outflows are launched and accelerated, how the different gas phases are coupled, how energy and momentum are distributed throughout the flow, and how the interaction between the wind and the host galaxy is ultimately realized \citep{Nardini+2015,Fiore+2017,Bischetti2019,Veilleux+2020, Laha2021}.
Addressing these questions requires a multi-wavelength approach capable of simultaneously probing the different phases of the outflow across cosmic time.

Two major observational challenges currently limit our understanding of the phenomenon. 
First, a comprehensive characterization of these winds demands a self-consistent combination of different tracers, complicating the construction of a unified picture \citep[e.g., ][]{Xu+2020, Fluetsch+2021}. These tracers include both emission and absorption lines that probe different gas phases across the entire electromagnetic spectrum, from the X-rays to the radio bands.
Second, the observability of individual tracers changes with redshift, as spectral features shift into different wavelength regimes and therefore become accessible to different instruments and observational techniques. This introduces significant heterogeneity in the available datasets, complicating the construction of uniform samples and limiting robust statistical comparisons across cosmic time.

Concerning the first point, it remains unclear whether emission- and absorption-based diagnostics measured in the same quasar spectrum can be physically associated with the same outflowing material.
In fact, emission and absorption tracers are generally sensitive to different gas conditions, as the former scales with the square of the gas density and the latter linearly with density, tracing denser and more diffuse environments, respectively.
Establishing a direct connection between the two is observationally challenging, mainly because their analyses often require different data from multiple instruments.
As a result, only a limited number of systems have been studied using both approaches simultaneously \citep[e.g.,][]{Bischetti+2024, Sharma+2025}.
In this context, the \civ$\lambda 1550$~\AA\ line provides a particularly valuable diagnostic, because it can be observed simultaneously in both emission and absorption within the rest-frame UV spectrum of quasars. 
In emission, \civ traces highly ionized nuclear winds through asymmetric and blue-shifted broad-line profiles \citep[e.g.,][]{marzianiMostPowerfulQuasar2016, Vietri+2018}.
In absorption, it probes outflows on larger galactic scales of tens to hundreds of parsecs, in the form of either narrow (NALs) or broad (BALs) absorption lines against the quasar continuum \citep{Arav+2012, Perrotta+2016, Perrotta+2018, He+2022}.
Despite studies reporting strong correlations between these phenomena \citep[e.g.,][]{Brazzini+2025}, the physical relation between \civ emission- and absorption-based outflow signatures remains debated. 
The picture becomes even more complex when attempting to connect \civ-based diagnostics with other commonly used outflow tracers, such as 
\oiii$\lambda 5007$\AA\ \citep[][]{Xu+2020,Vietri+2018} or \cii$158\mu$m \citep{Bischetti+2024, Zhu+2025} emission, which are used to trace ionized and cold neutral and molecular winds, respectively, on galactic scales of kiloparsecs. 

The second challenge is the evolution of outflow properties with redshift. 
Understanding how quasar-driven winds evolve across cosmic time is essential to constrain the growth of SMBHs, particularly in the early Universe when a substantial fraction of their mass was assembled through rapid and efficient accretion processes \citep{Inayoshi2020}. 
%High-redshift quasars are expected to differ significantly from their local counterparts, but a quantitative characterization of these differences is still an active area of investigation.
% Although the accretion physics is the same in both high-z and low-z quasars, the large-scale environments are not, and this is in turn expected to play a role in the properties of these objects: e.g. higher accretion rates in the early Universe, faster winds...
%OPTION 1
%In this work, we address both questions by investigating black-hole driven outflows in luminous ($\rm \log L_{bol}[ erg \ s^{-1}]>47$) quasars spanning from the Epoch of Reionization ($z\simeq 6-7$) to Cosmic Noon ($z\simeq 2-4$). We combine \civ emission and absorption line diagnostics to characterize the impact of feedback and the evolution of outflow properties across cosmic time.
% OPTION 2
In this work, we address both questions by investigating black hole-driven outflows, traced by the \civ transition in both emission and absorption, in luminous quasars ($\rm \log L_{bol}[ erg \ s^{-1}]>47$) spanning from the Epoch of Reionization ($z \simeq 6-7$) to Cosmic Noon ($z \simeq 2-4$). This enables us to characterize the impact of AGN feedback and the evolution of outflow properties across cosmic time.
Our sample encompasses the E-XQR-30 \citep{Dodorico2023-XQR30, Mazzucchelli+2023, Bischetti_2022, Bischetti_2023}, Qz5 \citep{Brazzini+2025} and SDSS datasets \citep{Shen+2011}, presented in Section \ref{sec:samples} and Figure \ref{fig:logLbol-vs-z}.
%The high-redshift sample consists of deep spectra of luminous, massive quasars accreting near or above the Eddington limit, including sources from the E-XQR-30 \citealp{Dodorico2023-XQR30, Mazzucchelli+2023, Bischetti_2022, Bischetti_2023} and HYPERION \citealp{Zappacosta+2023, Tripodi+2024} surveys. The low-redshift comparison sample is selected and uniformly reprocessed from the SDSS DR7 catalogue \citep{Shen+2011}. 
%All datasets are presented in Section \ref{sec:samples}.
In Section \ref{sec:spectral_modeling}, we describe the spectral fitting procedure adopted to characterize \civ emission and absorption lines, while in Section \ref{sec:analysis} we introduce the \civ-based diagnostics used to trace ionized outflows.
Finally, in Section \ref{sec:results} we present the results of our analysis, discussing the observed correlations between absorption- and emission-based outflow velocities, and their evolution with redshift.
%Finally, in Section \ref{sec:results} we present the results of our analysis, discussing how the observed correlations between minimum and maximum BAL velocities, as well as between BAL- and emission-based velocities, support a clumpy outflow structure and a common origin for BAL and nuclear winds, with the former likely arising from accelerated BLR gas clouds. 

Throughout the paper, we assume a $\Lambda$CDM cosmology with $H_0 = 67.3$ km s$^{-1}$ Mpc$^{-1}$, $\Omega _\Lambda = 0.68$, and $\Omega _M = 0.32$ \citep{planck_collaboration_planck_2020}. 
%%%%%%%%%%%%%%%%%%%%%%%%%%%%%%%%%%%%%%%%%%%%%%%%%%%%%%%%%%%%%%
\section{Quasar Samples}
\label{sec:samples}

\subsection{The high-redshift sample}

\begin{figure}
    \centering
    \includegraphics[width=\linewidth]{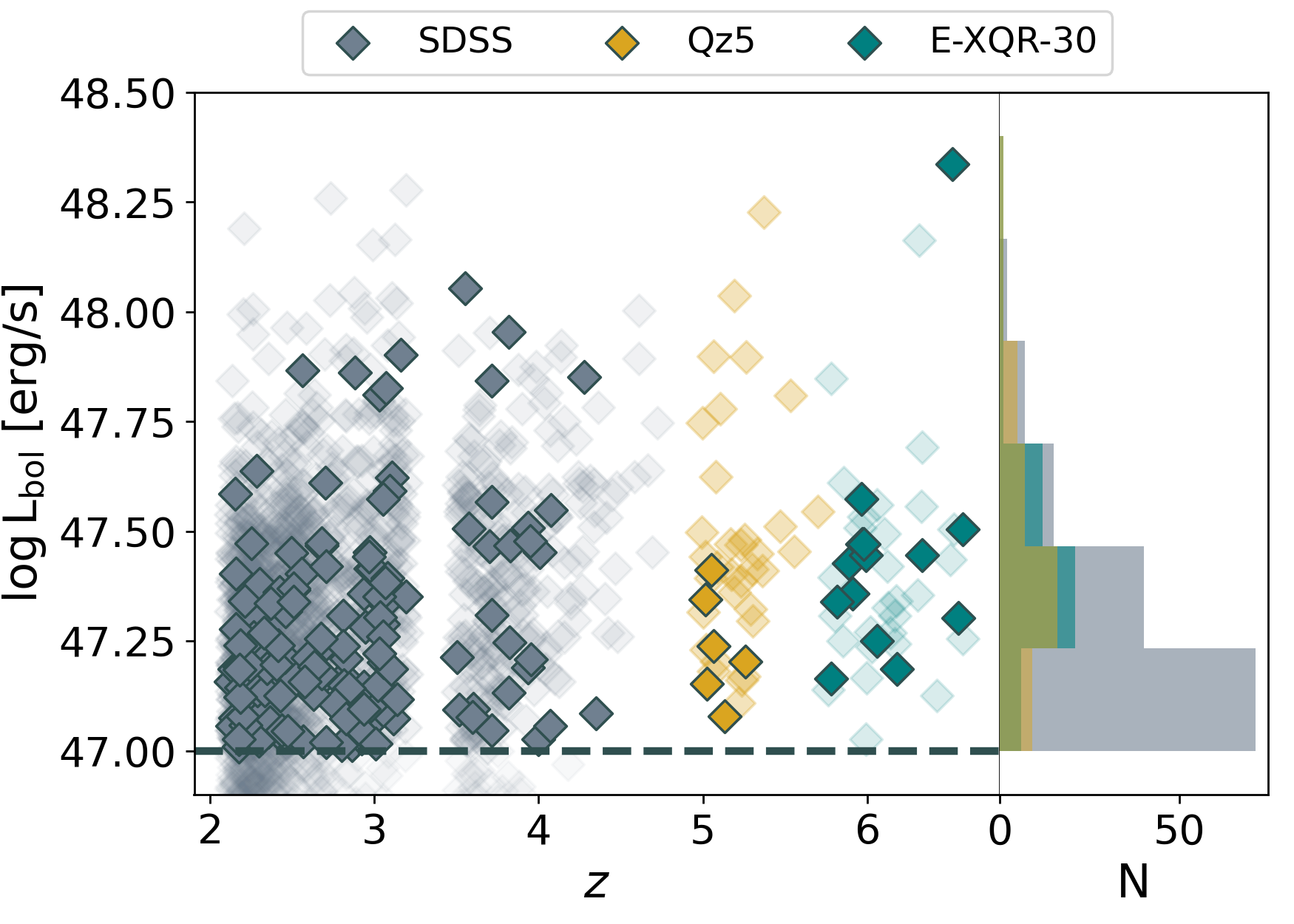}
    \caption{Bolometric luminosity versus redshift for E-XQR-30, Qz5 and SDSS quasar samples used in this work (semi-transparent symbols). 
    BAL quasars with $\log \rm L_{\rm bol} \ [erg\ s^{-1}] > 47$ (gray dashed line) and $\rm BI > 500$~km s$^{-1}$ are highlighted with full opacity. 
    The histogram on the right shows the $\rm L_{bol}$ distribution of all quasars satisfying the luminosity cut.
    The combined datasets span a wide redshift range, while probing only the high-luminosity end of the quasar population across cosmic time.
    %The right histogram represents the number of quasars per bin of 0.2 dex in $\rm \log L_{bol}$.
    }
    \label{fig:logLbol-vs-z}
\end{figure}

\begin{figure*}
    \centering
    \begin{subfigure}{\linewidth}
        \centering
        \includegraphics[width=0.9\linewidth]{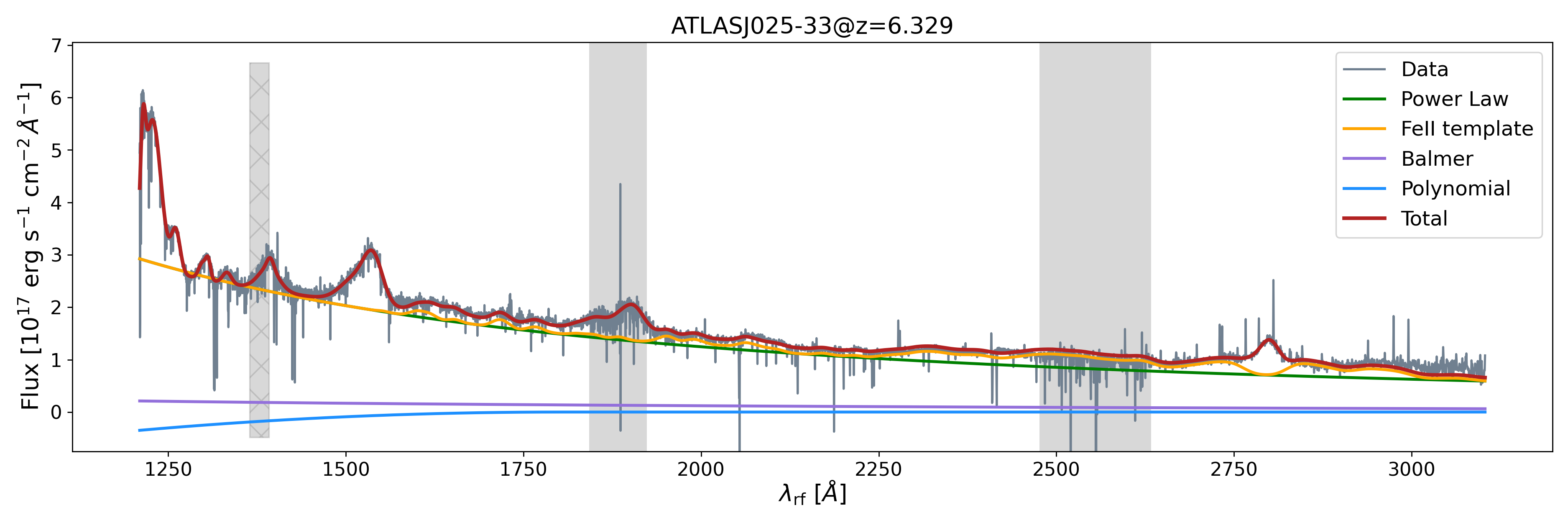}
    %\caption{Caption}
    \label{fig:quasar-spec1}
    \end{subfigure}
    \begin{subfigure}{\linewidth}
        \centering
        \includegraphics[width=0.9\linewidth]{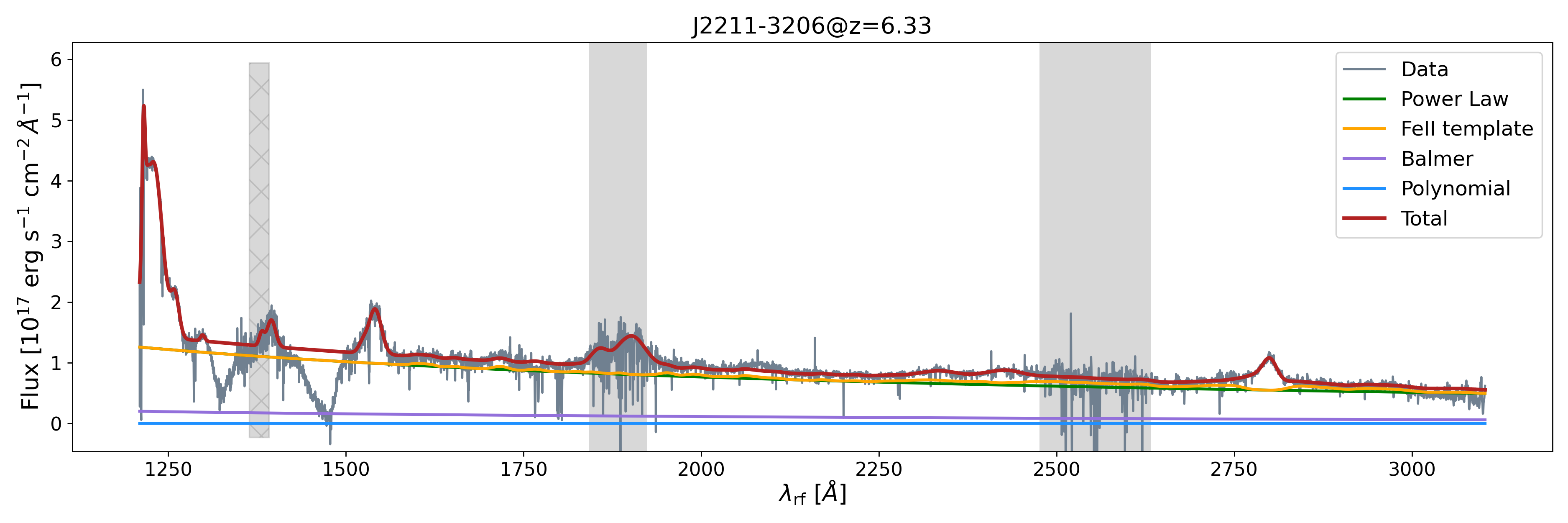}
    %\caption{Caption}
    \label{fig:quasar-spec2}
    \end{subfigure}
    \caption{Fits of continuum and line emission for the non-BAL quasar ATLASJ025-33 at $z=6.329$ (\textit{upper panel}) and for the BAL quasar J2211-3206 at $z=6.33$ (\textit{lower panel}). Spectra are reported in rest-frame, and the \lya forest is not displayed. 
    Cross-hatched gray regions indicate the overlap between the VIS and NIR arms of X-Shooter (occurring at observed wavelengths $\sim 1020$~nm), while gray-shaded regions highlight strong telluric contamination at observed wavelengths of $1350-1450$~nm and $1800-1950$~nm. 
    The total fit (continuum + emission lines) is shown in red, with the individual continuum components displayed singularly in different colours. The \feii emission below $1550$~\AA\ is set to zero (Section \ref{sec:spectral_modeling}).}
    \label{fig:quasar-spec}
\end{figure*}

\begin{figure*}
    \centering
    \includegraphics[width=\linewidth]{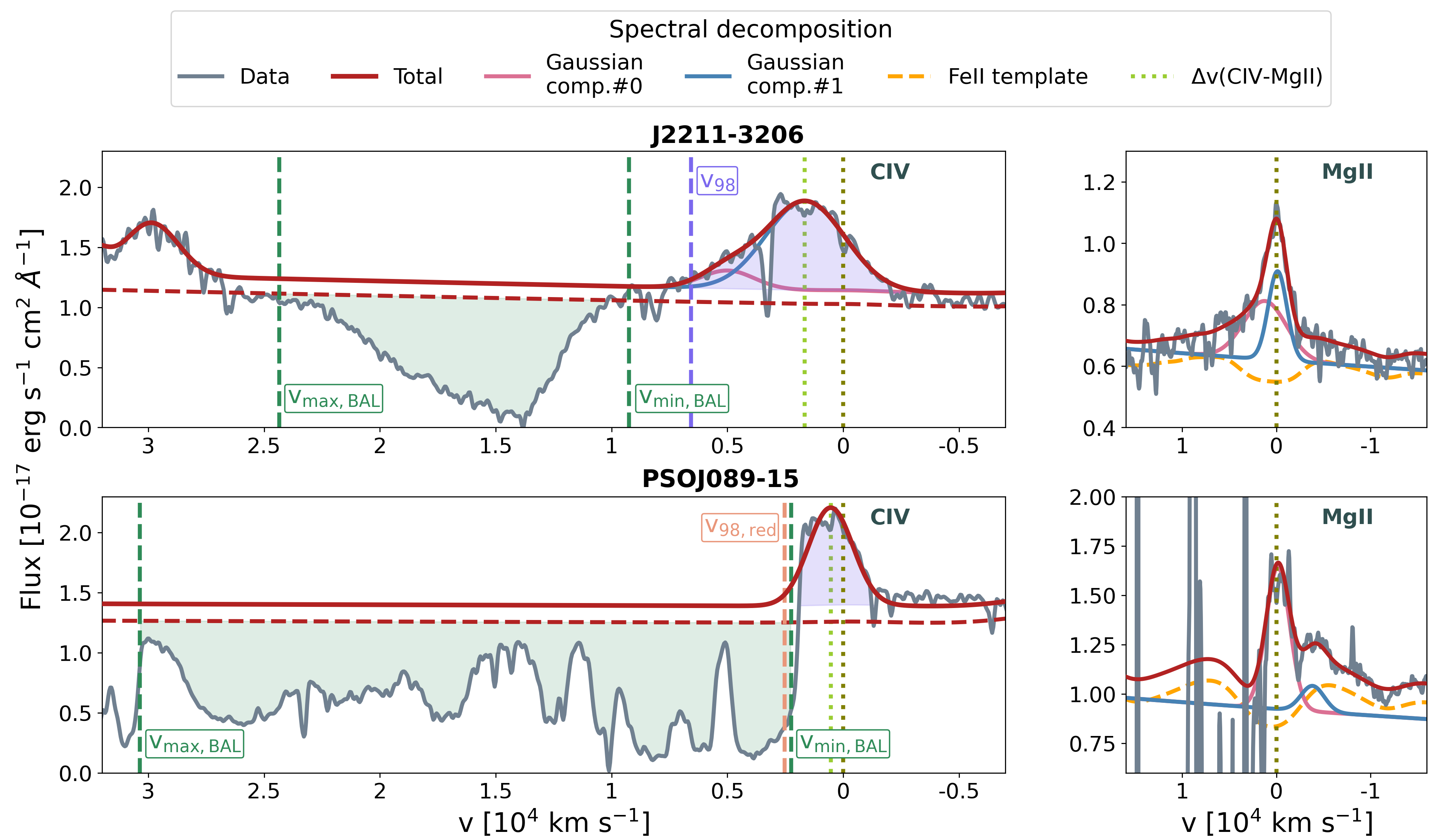}
    \caption{Zoom-in of spectral regions around \civ (\textit{left panels}) and \mgii (\textit{right panels}) emission lines for the BAL quasars J2211-3206 (see also Fig. \ref{fig:quasar-spec}) and PSOJ089-15 from E-XQR-30. 
    The total fit, including continuum and emission lines, is shown as a red solid curve. 
    Individual Gaussian components used to model each line are plotted in different colours, and in the right panels the \feii contribution is also shown as an orange dashed curve.
    The dark-green vertical dotted lines indicate the zero-velocity defined by the \mgii peak. 
    The \civ peak, which shows a clear blueshift in both quasars, is highlighted with a light-green dotted line. 
    \civ emission is filled in purple. 
    For J2211-3206 we retrieve $f>1$, therefore the entire \civ line profile is used to estimate \vel. In contrast, PSOJ089-15 has $f<1$, so we use \velred\ estimated from the red-side only.
    In the left panels, the red dashed line marks 90\% of the continuum level used to define the BAL trough (green-shaded area); \vminBAL\ and \vmaxBAL\ are taken from \cite{Bischetti_2022, Bischetti_2023}. 
    }
    \label{fig:emlines-example}
\end{figure*}

The Enlarged XQR-30 (E-XQR-30; \citealp{Dodorico2023-XQR30}) sample encompasses 42 high-redshift ($z=5.8 -6.6$), luminous quasars ($\log \rm L_{bol} [erg \ s^{-1}] = 47.0 - 48.3$; \citealp{Mazzucchelli+2023}, see also Table \ref{tab:bal_highz_info}) observed with the X-Shooter spectrograph \citep{vernet_x-shooter_2011} at the Very Large Telescope (VLT), and constitutes the largest homogeneous spectroscopic dataset of high-redshift quasars with high-quality rest-frame optical and UV observations. 
%, 30 of which belong to the Ultimate Legacy Survey of Quasars during Reionization (XQR-30; \citealp{Dodorico2023-XQR30}; see also \citealp{Mazzucchelli+2023}),
%\citep[XQR-30; ][]{Dodorico2023-XQR30}, 
%and 12 are selected from the X-Shooter archive (Figure \ref{fig:logLbol-vs-z}). 
%In this redshift regime, X-Shooter covers the rest-frame UV quasar emission, spanning from \lya to \mgii$\lambda 2799$\AA. 
The spectra have median resolving power of $R = 11400$, $9800$ in the VIS and NIR arms, respectively, and median signal-to-noise ratio (SNR) of $\sim 29$ per 10 km s$^{-1}$ pixel at rest-frame wavelength $1285$~\AA\ \citep{Bischetti_2022,Dodorico2023-XQR30}. 
The sample includes 17 BAL quasars \citep[][see also Section \ref{sec:absorption_line_identification}]{Bischetti_2022, Bischetti_2023}.

We complement E-XQR-30 with the Qz5 sample (\citealp{Brazzini+2025}; see also \citealp{Lai+2024, Oyarzun+2025, Wisz+2026}), which includes 39 luminous ($\log \rm L_{bol} [erg \ s^{-1}] = 47.0 - 48.2$) quasars at $z = 5.0-5.7$ observed with X-Shooter/VLT using the same instrumental setup as E-XQR-30. The Qz5 spectra have a median SNR of 13 per 10 km s$^{-1}$ pixel at $1285$~\AA. 
%using the same instrumental setup as E-XQR-30, allowing an extension of redshift coverage towards lower $z$. 
The sample includes six BAL quasars \citep{Brazzini+2025}. 
Originally designed for a blind search for damped Lyman-$\alpha$ systems (DLAs) at $z \sim 5$, it was selected to exclude quasars with previously known BAL systems and it is therefore not representative of the overall BAL quasar fraction at these redshifts.
%Consequently, it is not suitable for investigating the redshift evolution of the BAL fraction, as done in \cite{Bischetti_2023};  .

%\comm{Let's keep it here for now, then decide whether to insert this somewhere in a future discussion}
%The E-XQR-30 sample constitutes the largest homogeneous spectroscopic dataset of quasars at $z \gtrsim 6$ with high-quality rest-frame optical/UV observations. 
%In addition, many of these sources benefit from complementary multi-wavelength data, including ALMA \cii observations (venemans2016, Decarli2018, Neeleman+2021, \comm{who is writing the exqr30 alma paper? Maybe add some ref 'in prep'}), X-ray observations as part of the HYPerluminous quasars at the Epoch of ReionizatION (HYPERION) survey \citealp{Zappacosta+2023, Tripodi+2024}, and, more recently, also near-infrared imaging and spectroscopy from several JWST programs, including ASPIRE (refs), EIGER (refs), and AETHER (Farina+in prep). 

\subsection{The low-redshift sample}
We combine the high-redshift sample (E-XQR-30+Qz5; $z =5.0-6.6$) with the lower-redshift ($z = 2.1-5.0$) sample of 1905 quasars compiled by \cite{Bischetti_2022, Bischetti_2023} from the SDSS DR7 quasar catalogue \citep{Shen+2011}.
%to provide rest-frame optical coverage comparable to that of the high-redshift samples.
%This translates into 2MASS H and K bands detections for quasars with $z \lesssim 3.2$, and only 2MASS K detections for sources at $z>3.6$. 
To match the luminosity range of the higher-redshift dataset, we restrict to sources with $ \log \rm L_{bol} [erg \ s^{-1}] \geq 47$, resulting in a final sample of 1425 objects.
Differing from other works \citep[e.g., ][]{Laor+Brandt2002, Fiore+2017}, such a luminosity cut implies that here we focus on the high-luminosity tail of the quasar luminosity distribution (Fig. \ref{fig:logLbol-vs-z}).

\section{Quasar spectral modelling}
\label{sec:spectral_modeling}
We model the quasar rest-frame UV continuum and emission lines implementing the procedure detailed in \cite{Bischetti_2022} \citep[see also ][and references therein]{ Bischetti_2023, Mazzucchelli+2023}.
This method is consistently applied to E-XQR-30 and Qz5 samples.
%This method is consistently applied to the entire E-XQR-30 sample and to the six BAL quasars from Qz5, while the spectral analysis of the entire Qz5 sample is discussed in \cite{Brazzini+2025} (see also Appendix \ref{sec:appendix-exqr30-spectral-modeling}).
For SDSS, given the larger statistics, we adopt a more empirical procedure and do not perform analytical spectral fits (see Section \ref{sec:analysis}). 

\subsection{Continuum emission}
\label{sect:continuum_emission_modelling}

We model the quasar continua using spectral components that are fitted simultaneously over line-free spectral windows.  
These windows vary from source to source depending on their spectral characteristics, and typically include the rest-frame ranges $1270-1290$~\AA, $1650-1750$~\AA, $2000-2700$~\AA, and $2850-2950$~\AA. 
The intrinsic continuum emission is described by a power law corrected with a second-order polynomial \citep{Shen+2019}, to better reproduce the spectrum bluewards of \civ, particularly in reddened sources.
In practice, we perform two separate fits —one including the polynomial correction and one without— and adopt the model that minimizes the $\chi^2$ evaluated on the selected continuum windows. 
We further include a Balmer pseudo-continuum component, assumed to originate from partially optically thick hydrogen gas clouds with a uniform electron temperature of $\rm T_e = 15,000$~K \citep{Grandi1982, Dietrich+2002a, Dietrich+2002b}, and normalized such that its flux at $\lambda = 3675$~\AA\ equals 10\% of the power-law continuum at the same wavelength \citep[ e.g.,][]{Schindler+2020, Onoue+2020}. 
Finally, the \feii pseudo-continuum is modelled using the empirical template from \cite{Vestergaard+2001}, convolved with a Gaussian kernel 
%with a standard deviation ranging from $500$~km s$^{-1}$ to $5000$~km s$^{-1}$ 
to account for the velocity distribution of broad-line region (BLR) gas clouds in each source.
We do not consider \feii emission below $1550$~\AA, as \cite{Vestergaard+2001} caution that at shorter wavelengths their template is likely overestimated (see also Appendix \ref{sec:appendix-exqr30-spectral-modeling}).
Examples of continuum fits for a non-BAL and a BAL quasar are shown in Fig. \ref{fig:quasar-spec}, where the individual spectral components are highlighted in different colours. The fits for all E-XQR-30 quasars are reported in Figs. \ref{fig:all-quasar-spectra1}, \ref{fig:all-quasar-spectra2}. 

The most challenging region for continuum fitting is the one between \lya and \civ$\lambda 1550$~\AA, due to the low SNR at the junction of the VIS and NIR X-Shooter arms at an observed wavelength of $\sim 1020$~nm, and the presence of prominent UV lines (\siiv, \oiv, \nv) and multiple BAL troughs, whose identification is discussed in Section \ref{sec:absorption_line_identification}.  
In these cases, we assume that the observed continuum at $1270-1290$~\AA\ is representative of the intrinsic continuum emission.

%The same continuum fitting procedure is applied to the six BAL quasars from Qz5, differing from \cite{Brazzini+2025} where local fits were performed. Instead, we do not perform any continuum fitting for the WISSH and SDSS samples, and simply refer to the reference papers. 

\subsection{Emission lines}
\label{sect:emission_lines_modelling}

\begin{figure*}[t]
    \centering
    \includegraphics[width=\linewidth]{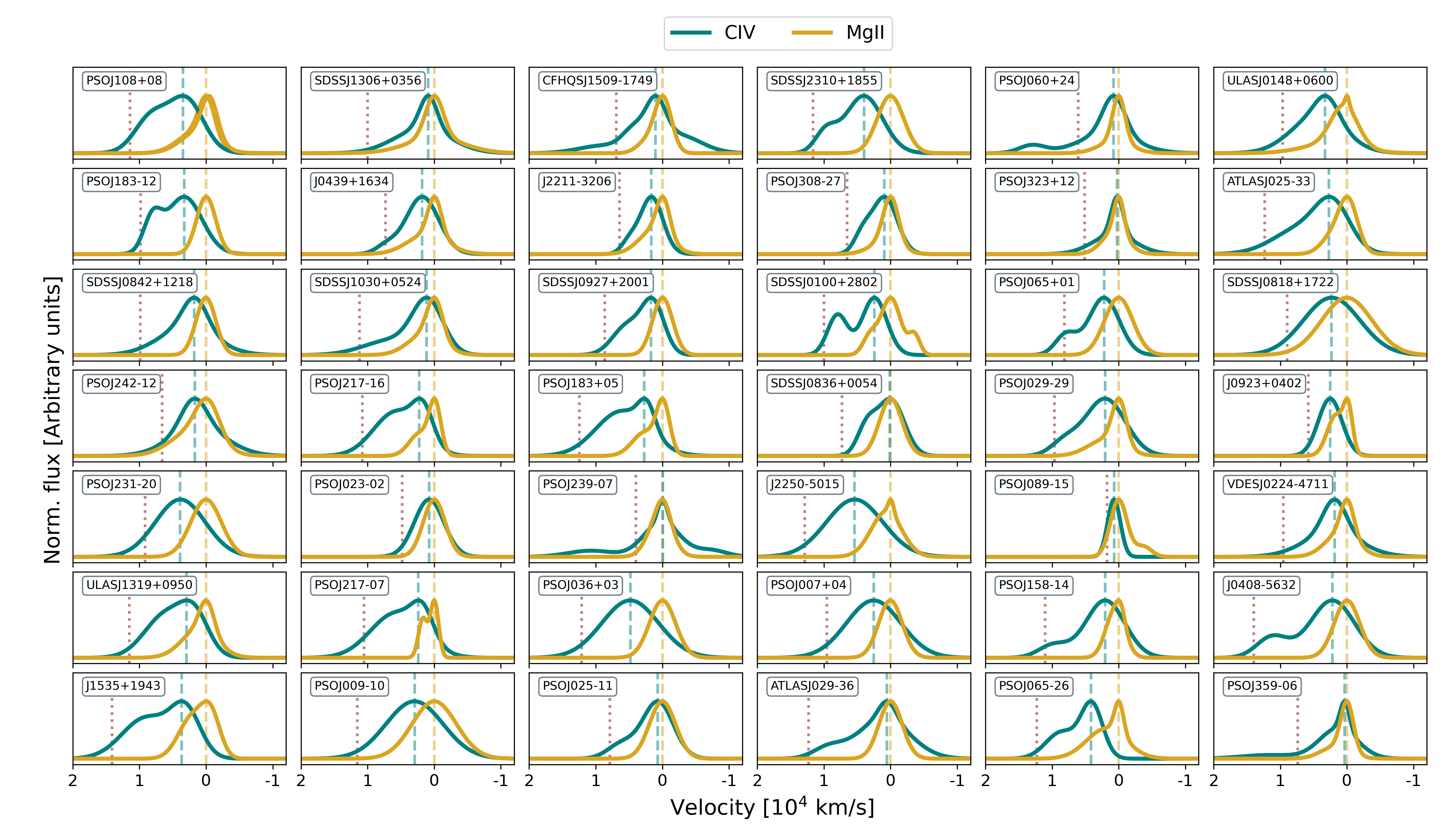}
    \caption{\civ (teal) and \mgii (gold) line model comparison for the E-XQR-30 quasars. The \mgii redshift is adopted as systemic (golden dashed vertical line), and line profiles are normalized to their peak to facilitate comparison. The teal dashed vertical line represents \deltav(\civ-\mgii), and the maroon dotted vertical line is \civ \vel. 
    %Most extreme blueshifts: the BAL J2250-5015 (-5400 km s$^{-1}$), the non-BAL PSOJ036+03 (-4820 km s$^{-1}$)
    }
    \label{fig:line-profiles-velspace}
\end{figure*}

We model emission-line profiles in the rest-frame, continuum-subtracted spectra using a multi-component Gaussian approach.
After fitting the continuum, the best-fit model is subtracted from the data, absorption features are masked, and the residual emission lines are fitted with one to three Gaussian components each.
Spectral regions affected by absorption, including BAL troughs, are automatically identified using a smoothed version of the spectrum as an approximate fit estimate. Pixels falling below 80-90\% of the smoothed spectrum are flagged as potential absorptions and excluded from the fit. The exact threshold is manually tuned depending on the SNR of each spectrum. This procedure is iterated twice. 
Emission-line parameters are then determined through non-linear least-squares minimization. Each Gaussian component is characterized by three free parameters: amplitude, centroid wavelength, and width. No kinematic constraints are imposed on these parameters, except that the line width is required to be $\leq 10^4$~km s$^{-1}$ to ensure a physically meaningful and well-constrained solution.

We focus in particular on \civ and \mgii lines, for which we adopt a baseline model consisting of two Gaussian components, representing systemic and outflow contributions. A third component is introduced when necessary to better reproduce the observed line profile, and is retained only if it contributes more than 10\% of the total line flux. 
In cases where SNR is low -- particularly for \mgii, where the SNR can drop to $< 10$ due to telluric absorption contamination in sources at $z \lesssim 6.0$ -- we restrict the fit to a single Gaussian component, as more complex models are not statistically justified.

The quasar emission redshift adopted for both continuum and emission-line fitting is derived from the \mgii line peak.
Starting from literature estimates \citep{Mazzucchelli+2023}, we iteratively repeat the continuum and emission-line fitting procedures, updating the redshift at each step until convergence to a self-consistent solution is achieved.

Two examples of our fitting results are shown in Fig. \ref{fig:emlines-example} for the BAL quasars J2211-3206 and PSOJ089-15, where the different Gaussian components are highlighted with distinct colours. 
The \civ and \mgii line profiles in velocity space for the entire E-XQR-30 sample are reported in Fig. \ref{fig:line-profiles-velspace}, revealing strikingly different shapes. More details on emission line properties are reported in Section \ref{sec:analysis} and in Appendix \ref{sec:appendix-exqr30-spectral-modeling}. 

\subsection{Absorption lines}
\label{sec:absorption_line_identification}

We adopt the balnicity index\footnote{We adopt the balnicity index definition reported in \cite{Gibson+2009}: \begin{equation}
    \text{BI} = \int_0^{v_{lim}} \bigg( 1 - \frac{f(v)}{0.9} \bigg) \cdot C\cdot  dv,
\end{equation}
where $f(v)$ is the continuum-normalized quasar spectrum, and the dimensionless parameter $C$ is set to 1 if $f(v) < 0.9$ in contiguous troughs with width $ \geq 2000$~km s$^{-1}$, and $0$ otherwise.} (BI) and the minimum and maximum BAL velocities, \vminBAL\ and \vmaxBAL, from \cite{Bischetti_2022, Bischetti_2023} for the E-XQR-30 and SDSS samples, updated with the new redshift estimates reported in this work, and from \cite{Brazzini+2025} for the six BAL quasars in the Qz5 sample. 
%For the E-XQR-30 sample, we revise the \vminBAL\ and \vmaxBAL\ measurements to account for the updated redshift estimates reported in this work.
In these studies, \civ BAL features are identified in the spectral region between \civ and \lya emission lines, allowing the exploration of BAL outflows with velocities up to $\rm v_{lim} \sim 64,000$~km s$^{-1}$.
The identification of \civ BAL troughs relies on the fact that the optical depth of \civ is usually comparable to, or larger than, that of \siiv \citep{Dunn+2012}, so that absorption features detected bluewards of \siiv are attributed to a low-velocity \siiv BAL only if a \civ BAL with similar velocity is observed. Otherwise, the absorption is considered to be produced by a high-velocity ($\rm v \gtrsim 30,000$~km s$^{-1}$) \civ BAL. 
This approach differs from earlier works \citep[e.g., ][]{Trump+2006, Gibson+2009}, which restricted the BAL search to wavelengths redwards of \siiv emission line, corresponding to $\rm v_{lim} \sim 25,000$~km s$^{-1}$. 
%However, in recent years more and more works have been pushing $\rm v_{lim}$ towards higher values, discovering that \civ BAL outflows with velocities exceeding $30,000$~km s$^{-1}$ are indeed more common than previously thought, at least in the high-redhsift regime (Cosmic Noon and beyond; see e.g. 
In particular, in these works BAL properties are derived using an empirical reconstruction of the quasar continuum, based on a composite template built from a large sample of non-BAL SDSS quasar spectra, selected to match the target spectrum within $\pm 20 \%$ in continuum colours $-$ expressed in terms of F(1700\AA)/F(2100\AA) and F(1290\AA)/F(1700\AA) flux ratios $-$ and equivalent width of \civ emission line. 
%The composite template is then normalized to the median flux value of the quasar spectrum in the rest-frame $1650–1750$\AA\ spectral interval, 
Further details of the method are provided in \cite{Bischetti_2023}. 
This approach enables a robust determination of BAL parameters, as it avoids explicit analytical continuum fitting and associated modelling uncertainties.
We therefore adopt this empirical method to derive the BAL parameters, rather than the analytical fits described in the previous section, since those fits are optimized for emission-line analysis rather than BAL measurements.
%After verifying that BAL parameters obtained with our analytical continuum from Section \ref{sec:spectral_modeling} are consistent with these empirical estimates \comm{EXPAND?}, we adopt the latter for consistency across the full sample, since the SDSS spectra are treated exclusively with this empirical method.

For the full sample, we impose a cut in balnicity index of $\rm BI > 500$~km s$^{-1}$, to select only stronger BAL troughs whose identification is unambiguous. 
The E-XQR-30 dataset counts 18 BAL quasars with BI spanning from $160$ to almost $14,000$~km s$^{-1}$, of which 14 have $\rm BI > 500$~km s$^{-1}$. 
All the six BAL quasars from Qz5 exhibit $\rm BI > 500$~km s$^{-1}$ spanning from $890$ to $\sim 6000$~km s$^{-1}$. The lack of extreme values likely derives from selection bias. 
For SDSS, this cut leads to a final sample composed of 187 BAL quasars. 
%The WISSH sample encompasses only two sources, J1157+2724 and J1549+1245. 

\section{Analysis of outflow tracers}
\label{sec:analysis}

We derive key emission-line properties for the analysed sample, focusing on the \civ-\mgii velocity shift, \deltav, and \civ \vel\ (Table \ref{tab:bal_highz_info}). 
% and \vmax (?)
The \civ-\mgii shift is defined as the velocity of the \civ line peak relative to the \mgii-based systemic redshift, with positive and negative values indicating blueshifts and redshifts of the \civ line, respectively.
%assuming velocities increase towards the left, so that positive and negative values indicate blueshifts and redshifts of the \civ line, respectively. 
The \vel\ parameter is defined as the velocity enclosing 98\% of the total emission line flux, computed by integrating the line profile from the red to the blue side. 
\vel\ is often used as a non-parametric, model-independent measure of the maximum velocity of outflows.
For E-XQR-30 and Qz5, where analytic emission-line fits are also available (Sect. \ref{sect:emission_lines_modelling}), we verify that the \vel\ values obtained from data integration are consistent with those from the best fit analytic profiles, and adopt the former throughout the analysis to ensure consistency with the SDSS measurements. 
%The latter is instead derived from line model and defined as \vmax$=|\Delta v|_{\rm out} + 2 \sigma_{\rm out}$, with $|\Delta v|_{\rm out}$ being the velocity shift between outflow and core components, and $\sigma_{\rm out}$ the velocity dispersion of the outflow component.
%Both quantities provide a measure of the maximum velocity of the outflowing gas, with \vel\ offering a model-independent estimate and \vmax\ relying on the decomposition of the line profile.
To account for the spectral noise, which is larger in the SDSS sample, we adopt a $+ 1 \sigma$ integration threshold above the continuum level, where $\sigma$ is the median continuum spectral error evaluated in spectral windows around \civ. 

%For the E-XQR-30 and Qz5 samples, this approach lowers \vel\ by $\sim 300$~km s$^{-1}$ on average, but the values remain consistent within $2 \sigma$ given the typical statistical error of $\sim 250$~km s$^{-1}$.
%The $+ 1 \sigma$ threshold is particularly useful for the SDSS sample analysis, where spectra are noisier and continuum estimates rely on an automated procedure rather than individual source analysis.

To deal with cases where the \civ line profile is strongly absorbed by a BAL system and \vel\ cannot be robustly measured directly on the blue side of the line, we introduce the parameter $f$, defined as the ratio between the total \civ line flux and twice the red-side only flux.
%, i.e. the flux derived by integrating \civ line emission only redwards of the line peak (purple-hatched region in Fig. \ref{fig:j2211-emlines}).
A symmetric \civ line yields $f=1$, while a prominent asymmetric blue wing results in $f \geq 1$; conversely, the presence of BAL features produces $f <1$. 
In the latter case (see e.g. Fig. \ref{fig:emlines-example}, lower panel, and Fig. \ref{fig:sdss-BALs-f_lower_1}), we adopt $\vel= \velred$, with \velred\ defined as the velocity enclosing 98\% of the red-side line flux, obtained by integrating the line profile from the line peak redwards.

Statistical errors on \deltav, \vel\ and \velred\ are estimated using Monte Carlo simulations. 
We generate N$=1000$ realizations of the \civ and \mgii spectral regions, assuming that the flux in each spectral bin follows a Gaussian distribution.
For each realization, we determine the line peaks and integrate the \civ line profile. The median and standard deviation of the resulting parameter distributions are adopted as the best-fit velocities and associated errors, respectively. 
For non-BAL quasars in the E-XQR-30 sample with $f>1$, we can achieve reliable estimates of both \vel\ and \velred, and compare them. 
We find that \vel\ is on average a factor 1.3 larger than \velred, indicating that \velred\ likely provides a conservative lower limit on \vel\ because it does not fully capture the blue-shifted \civ emission associated with outflows. 
To account for this effect, we adopt final asymmetric errors on \vel\ when $f<1$: the lower uncertainty is given by its statistical error derived from the Monte Carlo realizations, while the upper uncertainty is set to twice the average relative difference between \vel\ and \velred.

Lastly, we identify through visual inspection SDSS quasars for which \vel\ and/or \velred\ estimates cannot be robustly constrained because either the continuum and/or the \civ line peak cannot be identified, due to the low SNR of the spectra and/or the presence of BAL features (Fig. \ref{fig:sdss-BALs-excluded}). 
These sources are excluded from the subsequent analysis, resulting in a final SDSS BAL subsample of 129 quasars.

Our \deltav\ measurements for E-XQR-30 span from $\sim 0$~km s$^{-1}$ to more than $5000$~km s$^{-1}$, with a median value of $\sim 2100$~km s$^{-1}$\footnote{We cannot directly these results with those of \cite{Mazzucchelli+2023}, as in their work the \deltav\ is estimated with respect to the line centroids, rather than the line peaks.}. 
%They are consistent within $3 \sigma$ with those reported by \cite{Mazzucchelli+2023}, with a few exceptions likely due to differences in line decomposition and/or strong telluric contamination affecting \mgii.
The \vel\ distribution ranges from $2500$~km s$^{-1}$ to $14,000$~km s$^{-1}$, with a median of $9700$~km s$^{-1}$.
Restricting the analysis to BAL quasars does not significantly alter these results, yielding median \deltav\ and \vel\ of $2300$~km s$^{-1}$ and $9400$~km s$^{-1}$, respectively. 
In E-XQR-30, all selected BAL quasars with $\rm BI>500$~km s$^{-1}$ exhibit $f>1$, with the sole exception of PSOJ089-15, for which $f = 0.91$ (Fig. \ref{fig:EXQR30-BALs}), suggesting that \civ emission is partially absorbed by the BAL trough.
A peculiar source is PSOJ231-20, a BAL quasar with $\rm BI = 600$~km s$^{-1}$, which we exclude from our analysis because its unusual configuration places the \civ absorption entirely on the red side of \civ emission line. 
After applying the cut in balnicity index and excluding PSOJ231-20, the final E-XQR-30 BAL quasar sample comprises 13 sources.

The Qz5 sample covers similar ranges to E-XQR-30 in both \deltav\ and \vel, and all six identified BAL quasars exhibit $f > 1$ (Fig. \ref{fig:Qz5-BALs}). 
In contrast, the SDSS sample shows markedly different properties.
Both \deltav\ and \vel\ are systematically lower, with median values of $520$~km s$^{-1}$ and $6000$~km s$^{-1}$, respectively, when considering the full sample, and $600$~km s$^{-1}$ and $4000$~km s$^{-1}$ when restricting to BAL quasars, in agreement with the findings of \cite{Shen+2011} (see also \citealp{Richards+2002, Shen+2016}). 
The systematically lower \vel\ values measured for SDSS BAL quasars reflect the fact that the absorption trough often starts within the emission line, resulting in $f<1$.
In these cases, our measured \velred\ has to be regarded as a lower limit on true \vel, as discussed above. 
While in the high-redshift (E-XQR-30 + Qz5) BAL subsample only one quasar (out of 20 sources with BI$>500$~km s$^{-1}$) exhibits $f<1$, in the SDSS 74 out of 129 BAL quasars fall into this category. This implies that, at low redshift, more than 50\% of \civ BALs begin on the emission line, whereas at high redshift this fraction drops to only 5\%. As discussed more in detail in Section \ref{sec:redshift_evolution}, this difference may point to an evolution of BAL winds with redshift.

\section{Results}
\label{sec:results}

\begin{figure}[t]
    \centering
    \includegraphics[width=\linewidth]{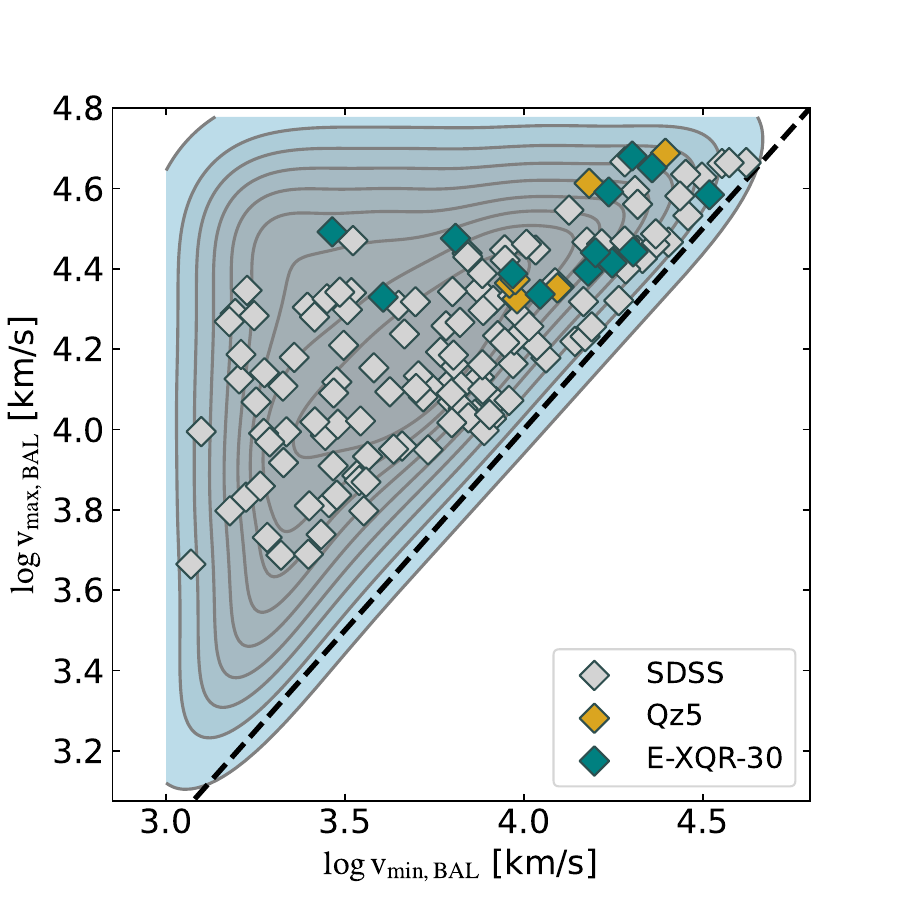}
    \caption{\vminBAL--\vmaxBAL\ distribution in logarithmic space. Observed data points from E-XQR-30, Qz5 and SDSS are reported in different colours. 
    %The high-redshift samples are systematically shifted towards higher BAL velocities. 
     The gray contour lines represent the probability distribution predicted by our clumpy outflow model after $10,000$ realisations, each containing $\rm N_c = 2$ clumps intersecting the line of sight (see Appendix \ref{Appendix:toymodel} for more details), with contours enclosing increasing fractions (from 10\% to 99.7\%) of the total model probability distribution. 
    The left and top boundaries correspond to the model minimum and maximum wind velocities ($v_0=10^3$~km s$^{-1}$ and $\rm v_{\infty} = 6\times10^4$~km s$^{-1}$ = 0.2c, respectively). 
    %While this distribution qualitatively covers the observed data, we do overestimate the density at very low $\log_{10}v_{\rm{max},\rm{BAL}} < 3.5$\,km s$^{-1}$ and in the top left corner for [$\log_{10}v_{\rm{min},\rm{BAL}} < 3.5$\,km s$^{-1}$, $\log_{10}v_{\rm{max},\rm{BAL}} >4.25$].
    }
    \label{fig:toymodel_vel_dist}
\end{figure}

\begin{figure*}[htbp]
  \centering
  \includegraphics[width=\linewidth]{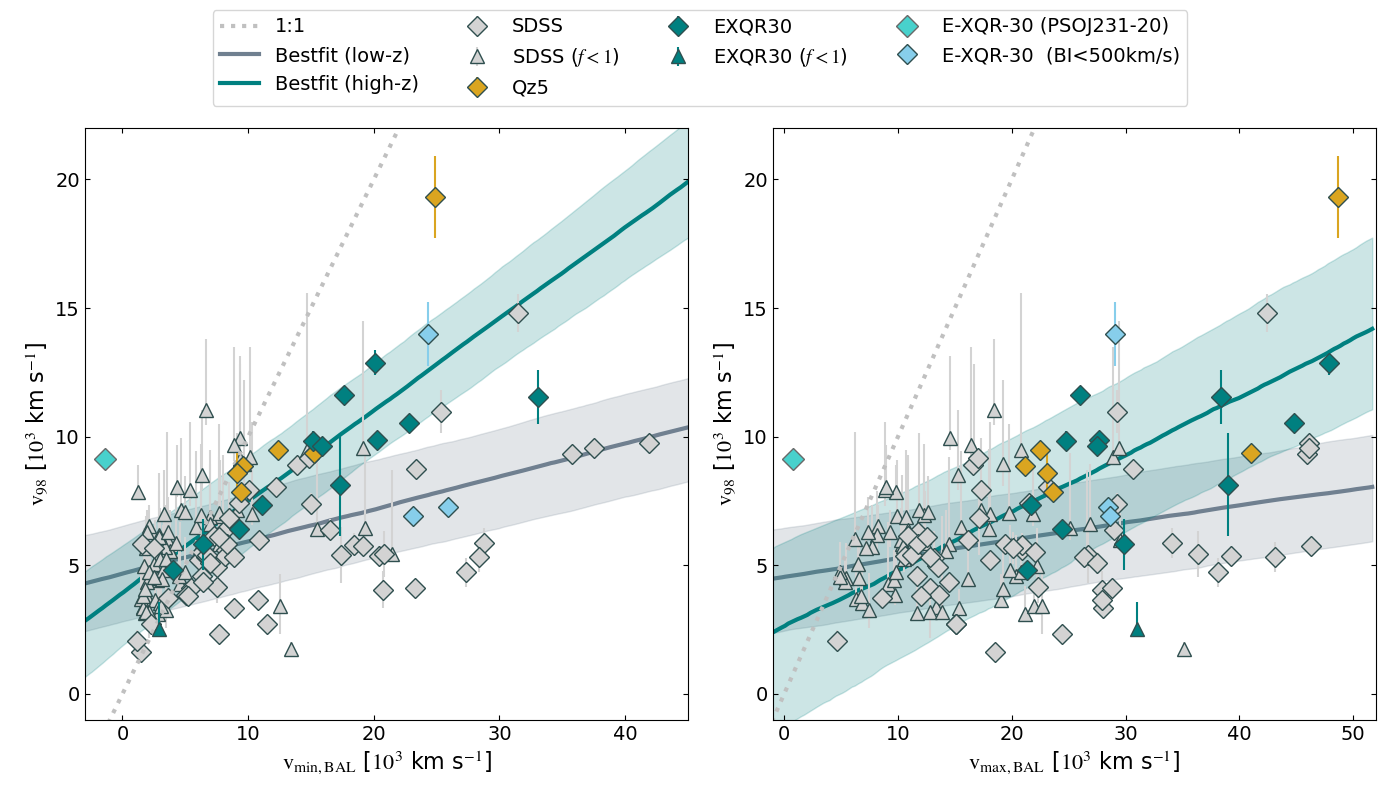}
  \caption{\civ \vminBAL--\vel\ (\textit{left panel}) and \vmaxBAL--\vel\ (\textit{right panel}) relations. 
  Data points are reported in different colours according to the original sample. 
  Data points with $f <1 $ (Sect. \ref{sec:analysis}) are shown as triangles. 
  For the E-XQR-30 sample, we further display the four BAL quasars excluded from our analysis: PSOJ231-20, and the three sources with $ \rm BI<500$~km s$^{-1}$ (see text for more details). 
  The best-fit relations for the high- and low-redshift samples are reported as a teal and gray solid line, respectively, and the $\pm 1 \sigma$ uncertainties as a shaded regions. }
  \label{fig:vel-relations}
\end{figure*}

\begin{table*}[]
\centering
\renewcommand{\arraystretch}{1.2}
\begin{tabular}{cccccccc}
\toprule
& Sample & Objects & $r$ & $p$-value & Slope & Intercept & Intrinsic scatter \\
& & & & & & [$10^3$ km s$^{-1}$] & [$10^3$ km  s$^{-1}$] \\
\midrule
\multirow{2}{*}{\vel-\vminBAL} & High-z & 13+6 & 0.840 & $6.70 \times 10^{-6}$ & $0.39^{+0.05}_{-0.04}$ & $4.12^{+0.56}_{-0.70}$ & $1.28^{+0.38}_{-0.29}$ \\
 & Low-z & 129 & 0.328 & $5.11 \times 10^{-5}$ & $0.11^{+0.02}_{-0.02}$ & $4.78^{+0.15}_{-0.20}$ & $1.89^{+0.13}_{-0.12}$ \\
\midrule
\multirow{2}{*}{\vel-\vmaxBAL} & High-z & 13+6 & 0.618 & $4.81 \times 10^{-3}$ & $0.23^{+0.07}_{-0.05}$ & $2.52^{+1.57}_{-1.95}$ & $2.61^{+0.61}_{-0.47}$ \\
 & Low-z & 129 & 0.287 & $4.15 \times 10^{-4}$ & $0.06^{+0.02}_{-0.02}$ & $4.62^{+0.24}_{-0.30}$ & $2.07^{+0.14}_{-0.13}$ \\
\bottomrule
\end{tabular}
\caption{Correlation statistics and best-fit parameters for the \vel-\vminBAL\ and \vel-\vmaxBAL\ relations considering separately the high- (E-XQR-30+Qz5) and low- (SDSS) redshift subsamples. In the first case, the correlations are statistically stronger, as suggested by the higher Pearson coefficients. The best-fit slopes are steeper, suggesting that, at a fixed \vel, high-redshift BAL quasars tend to have larger BAL velocities compared to their lower-redshift counterparts.}
\label{tab:statistics_of_vel_relations}
\end{table*}

\subsection{Acceleration of BAL winds}
\label{sec:Results:clumpyoutflows}

We compare \vminBAL\ and \vmaxBAL\ in Figure~\ref{fig:toymodel_vel_dist}. 
We find a strong correlation between the two BAL velocities, with a Pearson correlation coefficient, $r$, of $0.806$ and a p-value $p$ of $4.23 \times 10^{-35}$. 
This correlation remains significant even when including the SDSS quasars with low-velocity BALs, for which the \civ emission line is too absorbed to robustly measure \vel\ (Sec. \ref{sec:analysis}). 
A similar trend is also present when considering the high-redshift subsample alone (E-XQR-30 + Qz5; $r=0.650$, $p = 2.61 \times 10^{-3}$).
Interestingly, high-redshift quasars occupy a distinct high-velocity region in the \vminBAL-\vmaxBAL\ plane, and do not extend to the lower \vmaxBAL\ regime populated by SDSS quasars. This behaviour may indicate an evolution with redshift of quasar outflow properties, as we shall discuss more in detail in Section \ref{sec:redshift_evolution}. 

Based on the wind geometry proposed by \cite{Elvis2000}, we interpret the \vminBAL-\vmaxBAL\ relation within the framework of a clumpy outflow and develop a toy model to explain it, in which the BAL wind is composed of discrete clouds launched from the BLR and subsequently accelerated by radiation pressure (see Appendix \ref{Appendix:toymodel} for a detailed description). 
In this picture, the radial velocity increases with distance from the central black hole (Eq. \ref{eqn:v_l}), and \vminBAL\ and \vmaxBAL\ correspond to the velocities of the clumps intersecting the observer's line of sight closest to and furthest from the wind launching region, respectively.
The discrete nature of the outflow naturally introduces stochasticity in the observed BAL velocities, since the clump positions are randomly drawn from their underlying radial distribution (Eq. \ref{eqn:p_c}).

We assume the wind to cover a radial range of $\sim 1$--$6000$\,pc, consistent with the typical spatial scales inferred for BAL outflows \citep[e.g.][]{Arav+2012, Arav+2013, Arav+2018, Arav+2020_specific_single_source, Hemler+2019, Miller+2020, He+2022, Bischetti+2024}, a launch velocity of $\rm v_0 = 10^3\,\rm km\,s^{-1}$, and a terminal velocity $\rm v_{\infty} = 6\times10^4\,\rm km\,s^{-1} \simeq 0.2c$. 
%We assume inner and outer radial boundaries of $\rm R_0 = 10^4\,R_{\rm G}$ and $\rm R_{\rm max} = 10^6\,R_{\rm G}$, where $\rm R_{\rm G}=GM/c^2$. For a black hole mass of $\rm 10^{9.4}\,M_\odot$, corresponding to the average value of the E-XQR-30 sample, these limits translate into a radial range of $\sim 60$--$6000$\,pc, consistent with the typical spatial scales inferred for BAL outflows \citep[e.g.][]{Arav+2012, Arav+2013, Arav+2018, Arav+2020_specific_single_source, Hemler+2019, Miller+2020, He+2022, Bischetti+2024}. 
%We further adopt inner and terminal velocities of $\rm v_0 = 10^3\,\rm km\,s^{-1}$ and $\rm v_{\infty} = 6\times10^4\,\rm km\,s^{-1} \simeq 0.2c$, respectively. The lower limit reflects the fact that, by construction (see Section~\ref{sec-analysis}), our sample does not include sources with $\vminBAL \leq \vel$, while the upper limit is consistent with the maximum velocities expected for radiation-driven winds once relativistic effects are taken into account \citep{Luminari+2021}.
The full set of model parameters is presented in Appendix \ref{Appendix:toymodel}, together with the configuration providing the best qualitative agreement with the observations, shown in Fig.~\ref{fig:toymodel_vel_dist} as gray contours. 
In particular, the observed distribution of BAL velocities is best reproduced when, on average, $\rm N_c=2$ clumps intersect the observer's line of sight.
In this configuration, the model successfully captures the enhanced probability density just above the one-to-one relation, rather than at the extremes of the distribution.
A single clump ($\rm N_c=1$) naturally produces a one-to-one relation, while increasing the number of clumps systematically shifts the distribution towards more extreme values of both \vminBAL\ and \vmaxBAL.
This suggests a relatively low filling factor for the outflowing material along the line of sight, as independently retrieved in other literature works \citep[e.g.,][]{Arav+1994, Hamann+2013}.

This simplified model has several limitations. 
First, we adopt the same mass for all clumps, and the same velocity law, density profile, and number of clumps for all quasars. 
However, the quasars in our sample exhibit significantly different spectral energy distributions (SEDs), as expected given the $\sim2$ dex spread in black hole mass and Eddington ratio (Fig. \ref{fig:BHmass_LEdd_hist}). 
These differences in black hole properties are likely to impact the wind structure, including the velocity profile and mass outflow rate \citep[e.g][]{Proga+2000, Proga+2004}, particularly under the assumption of radiation-driven winds.
%For example, higher \LEdd\ produces stronger UV and softer X-ray emission \citep[e.g][]{KubotaDone2018, Mitchell+2023, Hagen+2024b}.
%\textcolor{red}{In some sense the effects of this can be seen in the top right corner of Fig.\,\ref{fig:toymodel_vel_dist}, where the E-XQR30 and Qz5 samples appear to preferentially occupy a region of parameter space with both high \vminBAL and high \vmaxBAL compared with the overall SDSS population. This could be explained due to the differences in the Eddington ratio distributions of the samples, shown in Fig.\,???, where the E=XQR30 and Qz5 distributions are significantly shifted towards higher $\lambda_{\rm{Edd}}$. $\lambda_{\rm{Edd}}$ significantly affects the spectral energy distribution (SED), generally giving stronger EUV emission and softer X-ray emission as $\lambda_{\rm{Edd}}$ increases \citep[e.g][]{KubotaDone2018, Mitchell+2023, Hagen+2024b}. If we interpret these winds as being due to radiation driving, then this change in SED shape and overall luminosity should have an impact on the wind velocity profile and mass outflow rate \citep[e.g][]{Proga+2000, Proga+2004}. Accounting for this in our toy model would require a more physical treatment of the winds, and is currently beyond the scope of the current paper. }
The second simplification is that the model assumes a fixed viewing angle through the outflow cone for all observers. In reality, a distribution of inclination angles is expected, introducing additional scatter through projection effects on the observed line-of-sight velocities.
Despite these simplifications, the model captures the main qualitative trends observed in the \vminBAL--\vmaxBAL\ plane, supporting the interpretation of BAL outflows as clumpy accelerating winds. In future work, we plan to extend this framework by incorporating more realistic wind geometries, distributions of physical properties, and orientation effects.

\subsection{Evidence of a common origin for BLR and BAL outflows}
\label{sec:Results:combining_abs+em}

We now investigate the connection between emission- and absorption-based outflow velocities, focusing on the \vel--\vminBAL\ and \vel--\vmaxBAL\ relations shown in Figure~\ref{fig:vel-relations}.
For the high-redshift sample, we find strong correlations between \vel\ and both BAL velocities, with the strongest relation observed for \vminBAL\ (Table~\ref{tab:statistics_of_vel_relations}).
Similar trends are present also when considering the low-redshift sample, although with lower statistical significance. 
A caveat here concerns SDSS quasars with low-velocity BALs, where absorption affects the \civ emission profiles. 
For objects in which the emission peak could still be identified, we estimate a lower limit on \vel\ using the red side of the line (\velred, as defined in Section \ref{sec:analysis}; triangles in fig. \ref{fig:vel-relations}). 
However, sources for which even the emission peak could not be reliably identified where excluded from the analysis. 
This selection can bias the \vel-\vminBAL\ correlation measured for the full sample, in the sense that, if we could recover the intrinsic \civ profiles of these excluded quasars, they would likely populate the low-\vminBAL\ region of the \vel-\vminBAL\ diagram above the 1:1 relation, further flattening the observed trend.
However, this effect does not impact the high-redshift sample, for which the \civ line peak can be robustly identified in all cases.
The correlation remains statistically significant even when combining the high- and low-redshift samples (\vel--\vminBAL: $r=0.548$, $p=5.82 \times 10^{-13}$; \vel--\vmaxBAL: $r=0.451$, $p=2.58 \times 10^{-5}$).
%As anticipated in Section \ref{sec:analysis}, this implies that relatively slow BAL winds absorbing the \civ emission line are common at low redshift, while at higher redshift, where winds are typically faster (Fig. \ref{fig:velocities_vs_z}), the line profile is less affected by BAL absorption. 

To quantify the \vel-\vminBAL\ and \vel-\vmaxBAL\ trends, we fit the observed relations with linear functions that also include an intrinsic scatter term $\sigma _{\rm int}$, such that the total variance of the $i$-th source is defined as $\sigma_{\mathrm{tot},i}^2 = \sigma_{\mathrm{meas},i}^2  + \sigma_{\rm int}^2$, where $\sigma_{\mathrm{meas},i}$ is the measured uncertainty, treated as asymmetric in cases where $f<1$. 
%The intrinsic scatter term is added after verifying that, if not included, we obtain best-fits with $\chi^2_{\rm red} \gg 1$.
The fits are performed within a Bayesian framework adopting non-informative flat priors and a Gaussian likelihood. 
We separately fit the low- and high-redshift samples (dark gray and teal solid lines in Figure~\ref{fig:vel-relations}, respectively). 
%We first fit the full dataset (maroon solid lines, with the shaded regions indicating the corresponding $\pm1\sigma$ confidence intervals), and then separately analyse the high-redshift and low-redshift (SDSS) subsamples.
The resulting best-fit parameters and correlation statistics are summarized in Table \ref{tab:statistics_of_vel_relations}.

The inferred intrinsic scatter remains always significant ($\sigma_{\rm int} \sim 1500$--$3000$ km\,s$^{-1}$) and is systematically larger in \vmaxBAL\ rather than \vminBAL, indicating that the observed relations are not fully deterministic and that additional physical processes likely contribute to the dispersion. Possible sources of scatter include differences in viewing angle, clump distribution within the flow, variability in the acceleration history, and projection effects associated with the outflow geometry.
Overall, the steeper slopes and stronger correlations of the high-redshift sample compared to the low-redshift sample suggest a different behaviour between high-z and low-z winds, as we shall discuss more quantitatively in Section \ref{sec:redshift_evolution}.

The observed \vel-\vminBAL\ and \vel-\vmaxBAL\ correlations suggest that the BLR gas clouds associated to the nuclear BLR winds, and traced with \civ \vel, are physically linked to the larger-scale BAL winds.
We therefore propose that BAL winds are composed, at least in part, of gas clouds originating from the BLR itself. Such a common origin would naturally explain the observed velocity correlations.
Moreover, the fact that \vminBAL\ usually exceeds \vel, particularly in the high-velocity regime, implies that the BAL outflow is observed after it has already undergone significant acceleration. In the context of our model, this suggests that the BAL wind originates (or, at least, becomes visible) at a radius $\rm R_{0,BAL}$ larger than the BLR typical radius, which corresponds to the launching radius of the outflow ($\rm R_{0,BAL} > R_0 \equiv R_{BLR}$; Figure \ref{fig:toymodel_geometry}).

\subsection{Redshift evolution of outflow velocities}
\label{sec:redshift_evolution}

\begin{figure}
    \centering
    \includegraphics[width=\linewidth]{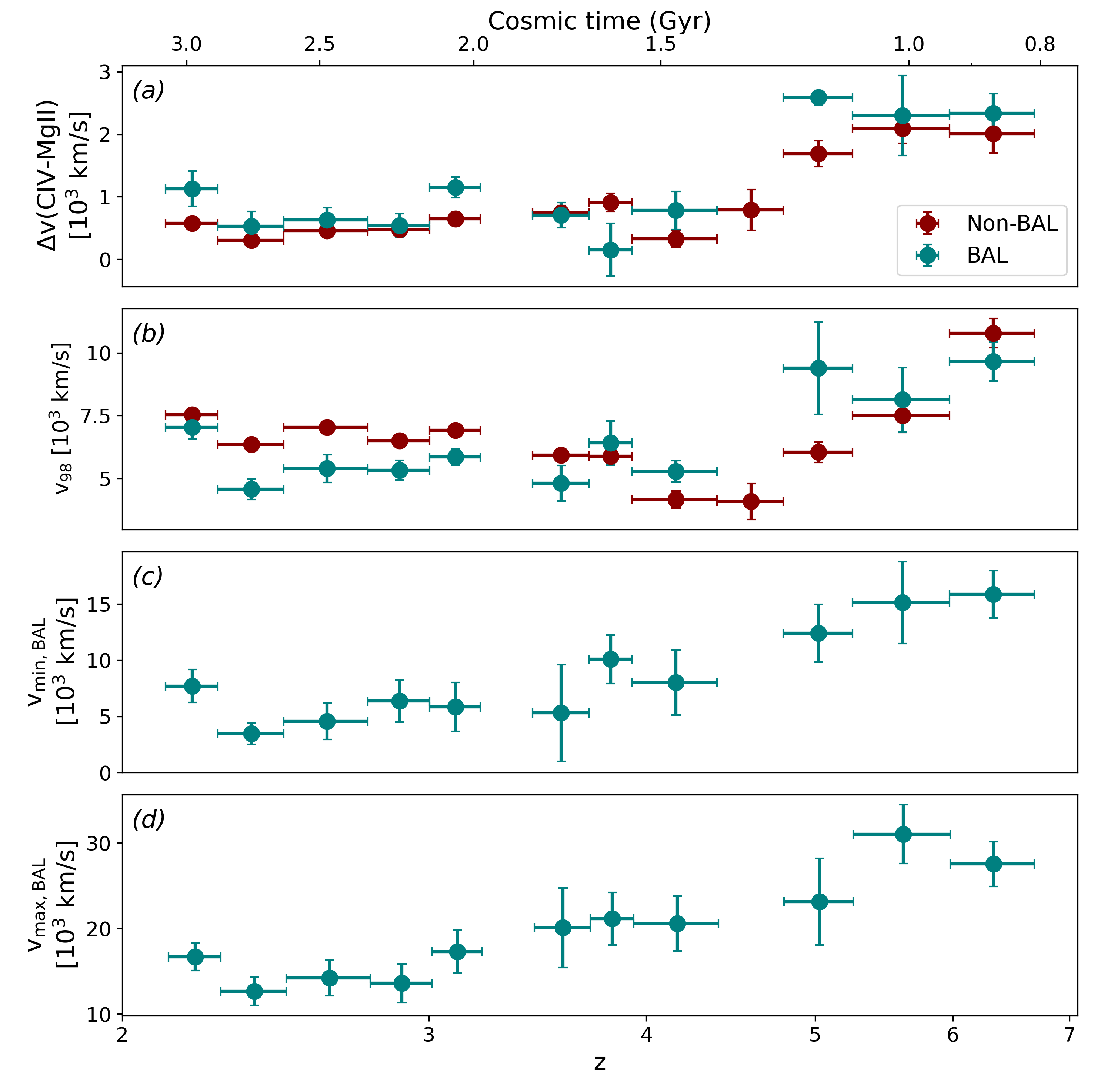}
    \caption{\civ-\mgii velocity shift (panel a), \vel\ (panel b), minimum \civ BAL velocity (panel c), and maximum \civ BAL velocity (panel d), as a function of redshift and cosmic time (top axis). Horizontal bars indicate the individual redshift bins in which the average velocities have been calculated ($\Delta z \sim 0.2$ in the lower redshift regime covered by the SDSS, and $\Delta z \sim 0.7$ at higher redshift, corresponding to bin widths in cosmic time of $\Delta t \sim 100-300$~Myr). Dark red points indicate the non-BAL quasar population satisfying the luminosity cut $\log \rm L_{bol} [erg \ s^{-1}]> 47$, while the teal points correspond to the BAL subsample with $\rm BI  > 500$~km s$^{-1}$. Vertical bars indicate the error associated to the median velocity in each bin. }
    \label{fig:velocities_vs_z}
\end{figure}

%The evolution with redshift of \deltav\ and, more recently, of \vminBAL\ and \vmaxBAL\ has been extensively studied in the literature \citep[e.g., ][]{Meyer+2019, Schindler+2020, Bischetti_2022, Bischetti_2023, Tortosa+2024}.
Visual inspection of Figures \ref{fig:toymodel_vel_dist} and \ref{fig:vel-relations} shows that high-redshift quasars occupy a distinct region of the parameter space compared to their lower-redshift counterparts, exhibiting systematically higher velocities from both emission- and absorption-based diagnostics.
This, in turn, points towards an evolution of outflow properties with redshift.
We investigate these evolutionary trends in Figure \ref{fig:velocities_vs_z}, which reports the binned values of \deltav, \vel, \vminBAL\ and \vmaxBAL\ as a function of $z$.
For the emission-based diagnostics, we display binned values calculated for both BAL and non-BAL subsamples.
The binning intervals are chosen to ensure comparable BAL quasar statistics across the full redshift range, adopting $\Delta z \sim 0.2$ in the lower redshift regime covered by the SDSS (where each bin encompasses 20 BAL quasars), and $\Delta z \sim 0.7$ at higher redshift (where each bin encompasses 5 to 7 BAL quasars). 
This corresponds to a cosmic time binning of $\Delta t \sim 100-300$~Myr.

Both \deltav\ and \vel\ remain approximately constant over the redshift range $z \sim 2 -4$, and then increase at $z \gtrsim 5$, with \deltav\ rising from a median value of $\sim 600$~km s$^{-1}$ to $\sim 2500$~km s$^{-1}$, and \vel\ from $\sim 6000$~km s$^{-1}$ to almost $10,000$~km s$^{-1}$. 
Evidence for an increase in emission-based outflow velocities in the early Universe has already been extensively investigated in the literature \citep[e.g., ][]{Meyer+2019, Schindler+2020, Tortosa+2024, Belladitta+2025}, and our results further confirm this trend. 
%For instance, the largest \deltav\ discovered so far is of almost $10^4$~km s$^{-1}$, presented in \cite{Belladitta+2025} for a quasar at $z=6.3$.
%To assess whether these evolutionary trends in emission velocities depend on the presence of BAL features, we perform a statistical two-sample Kolmogorov-Smirnov (KS) test comparing the \deltav\ and \vel\ distributions of BAL and non-BAL quasars. In both cases, we find no statistically significant difference between the two populations (KS statistic $D = 0.015, 0.025$ for \deltav\ and \vel, respectively, with corresponding p-values of $p=0.10, 0.07$). This suggests that emission-based outflow diagnostics do not depend on the presence of BALs, and that their redshift evolution is representative of the quasar population as a whole, consistent with the interpretation that the manifestation of BAL features is simply a matter of viewing angle. The statistic is slightly weaker for \vel, likely due to the limitations discussed in Section \ref{sec:analysis}, in particular the large number of \vel\ lower limits ($f<1$) encountered at low redshift.

A similar redshift evolution is observed for the BAL velocities, although in this case the trend appears smoother, consistent with the findings of \cite{Bischetti_2022, Bischetti_2023}, which we complement with the inclusion of the $z \sim 5$ regime\footnote{\cite{Bischetti_2023} adopt a different binning procedure, constant in cosmic time with $\Delta t \sim 200$~Myr, while our binning procedure is based on the optimization of the BAL statistics.}. 
%In particular, \cite{Bischetti_2023} adopt a different binning procedure, constant in cosmic time, with $\Delta t \sim 200$~Myr, while our binning procedure is based on the optimization of the BAL statistics. Nevertheless, we verify that the two different binning procedures provide consistent results. 
In the most extreme cases, we identify BAL outflows with maximum velocities reaching a significant fraction of the speed of light (up to $\sim 0.15$~c), such as J2250-5015 and PSOJ217-07 in E-XQR-30, or J1004+2025 in Qz5.
The discovery of high-redshift \civ BALs with such extreme velocities is becoming increasingly common, as new methods are developed to identify high-velocity \civ absorption also bluewards of the \siiv emission, overcoming the traditional limit of $\rm v_{lim}=25,000$~km s$^{-1}$ \citep[e.g., ][]{Bruni+2019, Rodriguez-Hidalgo+2020, Rodriguez-Hidalgo+2025, Wang+2021, Vietri+2022, Vietri+2025}. 
Recently, \cite{Seaton+2026} have reported the fastest \civ BAL outflows known to date in a luminous quasar at Cosmic Noon, with \vmaxBAL\ extending to at least $\sim 90,000$~km s$^{-1}$, corresponding to absorption bluewards of \lya. 
However, such extreme velocities are not accessible at higher redshifts, where the increasingly dense \lya forest makes it difficult to disentangle \civ absorption from blending with numerous intervening hydrogen absorption lines. If present, such extreme outflows would further strengthen our results, by increasing even more the median values of \vminBAL\ and \vmaxBAL\ in the higher-redshift bins.

Remarkably, all velocity diagnostics considered in this work (\deltav, \vel, \vminBAL\ and \vmaxBAL) show little evolution up to $z \sim4$, followed by a  significant increase at $z \gtrsim 5$. This suggests that $z \sim 5$ may represent a transition epoch between a phase of efficient black hole feedback in the early Universe and a successive, less extreme maintenance phase characteristic of lower redshifts.
To quantify the significance of this evolution, we compare the velocity distributions of high- and low-redshift subsamples using two statistical tests: the two-sample Kolmogorov-Smirnov (KS) test \citep{KStest}, and the Mann-Whitney (MW) test \citep{MWtest} U-test.
Both tests show that all the velocity distributions differ significantly ($> 4 \sigma $) between the two redshift regimes, with KS statistic exceeding 3 for all velocity diagnostics and associated p-values $p _{\rm KS} <3 \times 10^{-5}$, and MW p-values $p _{\rm MW} <2.8 \times 10^{-4}$. A similarly statistically robust difference is found also when considering only BAL quasars. 
To test whether this result depends on the exact redshift division, we repeat the analysis using different redshift thresholds ($z=4.5, 5 $ and $5.3$), rather than a division by samples. 
In all cases, the high- and low-redshift samples remain significantly different ($> 3\sigma$), supporting the conclusion that a change in outflow properties occurs around $z \sim 5$.
We next investigate whether this difference in outflow properties reflects underlying variations in black hole properties, such as black hole masses and/or accretion rates, while we do not test bolometric luminosity dependence, since the sample is already luminosity-matched (Sect. \ref{sec:samples}).
To this end, we estimate $\rm M_{BH}$ and \LEdd\ using both \civ and \mgii estimators, following \cite{VestergaardPeterson2006_CIV} (with the correction by \citealp{Coatman+2017}) and \cite{VestergaardOsmer2009_MgII}, respectively (see also Figure \ref{fig:BHmass_LEdd_hist}). 
%In fact, high-redshift quasars are known to accrete at higher Eddington ratios \LEdd\ than local sources (refs), and the same trend is present also in our sample, as we can see from Figure TODO (Ledd hist). 
In both cases, KS and MW tests reveal no statistically significant difference between \LEdd\ and $\rm M_{BH}$ distributions of high- and low-redshift quasars. 
Using \civ-based \LEdd\ estimates yields only marginal differences of less than $3 \sigma$, which become even less significant when using \mgii\footnote{The statistical tests based on \mgii encompass only the SDSS subsample for which this line is covered, that is, 332 quasars at $z \leq 2.25$.} ($ <2 \sigma$). 
Differences in black hole masses are even weaker.
We therefore conclude that the observed evolution of outflow properties with redshift is not driven by an evolution of black hole properties, in agreement with \cite{Bischetti_2023}, who found that BAL wind properties do not depend on \LEdd. 

\section{Conclusions}

In this work, we analyse a sample of luminous quasars ($\log \rm L_{bol} [erg \ s^{-1}]>47$) spanning from the Epoch of Reionization to Cosmic Noon, focusing on the high-luminosity tail of the quasar population (Fig. \ref{fig:logLbol-vs-z}).
Our sample combines the E-XQR-30 ($z \sim 6 - 6.6$), Qz5 ($z \sim 5 - 6$) and SDSS ($z \sim 2- 5$) datasets, which are luminosity-matched and homogeneously analysed.
We investigate ionized outflows through complementary emission- and absorption-based velocity diagnostics based on the \civ line, with the aim of studying (1) the connection between different observational tracers of quasar winds, and (2) the evolution of the inferred outflow properties across cosmic time. 
We adopt \deltav, \vel, and the minimum and maximum BAL velocities as emission- and absorption-based outflow diagnostics, respectively.
Our main findings can be summarised as follows: 
\begin{itemize}
    \item We find a statistically significant correlation between \vminBAL\ and \vmaxBAL\ (Fig. \ref{fig:toymodel_vel_dist}), which can be qualitatively reproduced using a toy model of accelerating clumpy outflows. This result supports a scenario where BAL winds consist, at least in significant part, of an inhomogeneous medium composed of dense gas clumps.
    \item We further identify statistically significant correlations between \vel\ and the BAL velocities, particularly \vminBAL\ (Fig. \ref{fig:vel-relations}, Table \ref{tab:statistics_of_vel_relations}). 
    These correlations indicate a close physical connection between emission- and absorption-traced winds. 
    Since \vel\ is commonly associated with BLR nuclear winds, we propose that BAL winds are composed, at least in part, of BLR gas clouds that are expelled and accelerated to larger scales. This is also consistent with the clumpy nature of the BAL outflow, as probed in the previous point.
    \item The fact that \vminBAL\ usually exceeds \vel\ suggests that BAL winds become observable only after the gas has undergone some acceleration relative to the characteristic BLR velocity. 
    \item High-redshift quasars occupy a distinct high-velocity region in the \vminBAL-\vmaxBAL\ diagram (Fig. \ref{fig:toymodel_vel_dist}), and exhibit a tighter and steeper correlation between \vel\ and the BAL velocities (Fig. \ref{fig:vel-relations}). This behaviour points towards an evolution of quasar outflow properties with redshift. 
    \item Both the emission-based (\deltav\ and \vel) and absorption-based (\vminBAL\ and \vmaxBAL) diagnostics  remain approximately constant up to $z \sim 4$, followed by a marked transition at $z \sim 5$ (Fig. \ref{fig:velocities_vs_z}). This suggests that $z \sim 5$ may represent a transitional epoch between an early phase of efficient black hole feedback and a successive, less extreme maintenance phase characteristic of lower redshifts.
    The distinction between high-redshift (E-XQR-30 + Qz5) and low-redshift (SDSS) quasar populations is quantitatively confirmed with statistical  Kolmogorov-Smirnov (KS) and Mann-Whitney (MW) U-tests.
    \item Additional KS and MW statistical tests show that  the observed evolution of outflow properties with redshift is not driven by an evolution of black hole properties, in agreement with \cite{Bischetti_2023}, who found that BAL wind properties do not depend on \LEdd.
     
\end{itemize}

This work highlights the importance of investigating ionized outflow properties in luminous quasars over the widest possible redshift range.
In the low-redshift regime, the major limitation is the relatively modest resolution and SNR of SDSS data. 
At high redshift, the situation remains poorly explored, particularly at $z \sim 5$, where the limited available samples are also biased against BAL quasars.  
Improving both the sampling of the high-redshift regime and the data quality at low redshift, using for example DESI and 4MOST surveys, will significantly increase the diagnostic power of this approach. 
However, no ongoing or planned surveys currently target the $z \sim 5$ regime in a systematic way, despite this redshift range being a key epoch of our work, as it probes the expected transition between the highly efficient feedback phase characteristic of the early Universe and a less extreme regime typical of lower redshifts.

%%%%%%%%%%%%%%%%%%%%%%%%%%%%%%%%%%%%%%%%%%%%%%%%%%%%%%%%%%%%%%
\begin{acknowledgements}
      The research activities described in this paper were carried out with contribution of the Next Generation EU funds within the National Recovery and Resilience Plan (PNRR), Mission 4 - Education and Research, Component 2 – From Research to Business (M4C2), Investment Line 3.1 – Strengthening and creation of Research Infrastructures, Project IR0000034 – “STILES - Strengthening the Italian Leadership in ELT and SKA”. 
      AT acknowledges support from the INAF Large Program “DELUX” of the “Ricerca Fondamentale 2024” INAF program.
      FS acknowledges financial support from the Bando Finanziamento ASI CI-UCO-DSR-2022-43 CUP:C93C25004260005 project ``IBISCO: feedback and obscuration in local AGN'', Ricerca Fondamentale INAF 2024 under project 1.05.24.07.01 MINI-GRANTS RSN1 "ECHOS". CF and FS acknowledge support from the Ricerca Fondamentale INAF 2023 Data Analysis grant 1.05.23.03.04 ``ARCHIE ARchive Cosmic HI \& ISM  Evolution''.
      LZ acknowledges support from the INAF Large Program “DELUX” of the “Ricerca Fondamentale 2024” INAF program.
\end{acknowledgements}

%%%%%%%%%%%%%%%%%%%%%%%%%%%%%%%%%%%%%%%%%%%%%%%%%%%%%%%%%%%%%%
% WARNING
% Please note that we have included the references below in
% order to compile the document, but we ask you to:
%
% - use BibTeX with the regular commands:
%   \bibliographystyle{aa} % style aa.bst
%   \bibliography{Yourfile} % your references Yourfile.bib
% - join the .bib files when you upload your source files
%%%%%%%%%%%%%%%%%%%%%%%%%%%%%%%%%%%%%%%%%%%%%%%%%%%%%%%%%%%%%%

\bibliographystyle{aa}
\bibliography{bib}

%%%%%%%%%%%%%%%%%%%%%%%%%%%%%%%%%%%%%%%%%%%%%%%%%%%%%%%%%%%%%%%
% Appendices must be placed after   \end{thebibliography}
% They will be placed automatically on a new page.
%%%%%%%%%%%%%%%%%%%%%%%%%%%%%%%%%%%%%%%%%%%%%%%%%%%%%%%%%%%%%%%
\begin{appendix}

\onecolumn

\section{E-XQR-30 and Qz5: Spectral modelling and derived properties}
\label{sec:appendix-exqr30-spectral-modeling}

\begin{figure*}[ht!]
    \centering
    \includegraphics[width=\linewidth, trim={0cm 0cm 0cm 0cm}]{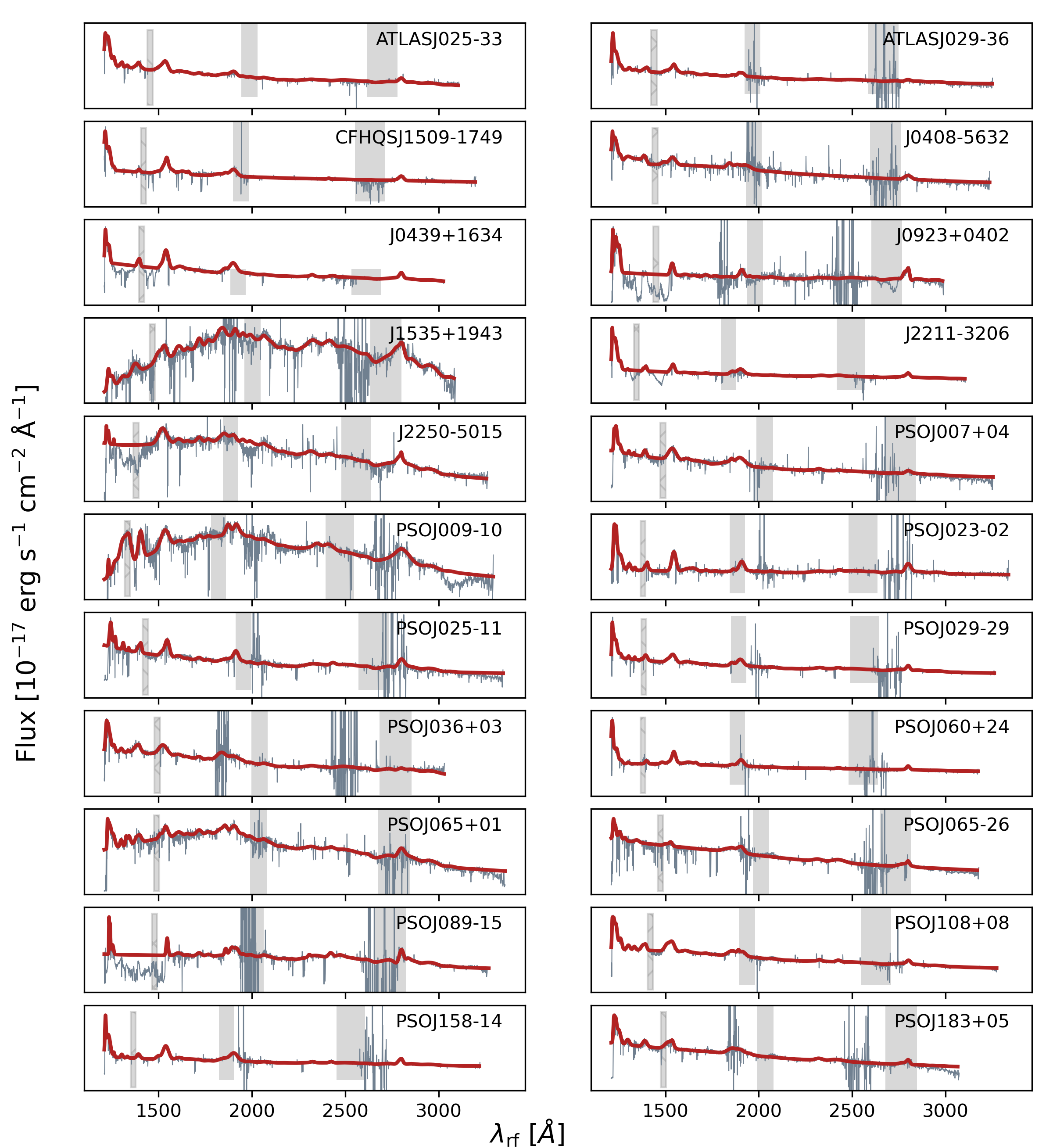}
    \caption{Rest-frame spectra and best-fit models for the E-XQR-30 quasars. The gray-shaded regions mark the overlap between the VIS and NIR arms, as well as spectral regions affected by telluric absorption.}
    \label{fig:all-quasar-spectra1}
\end{figure*}

\begin{figure*}[ht!]
    \centering
    \includegraphics[width=\linewidth, trim={0cm 1.7cm 0cm 0cm}]{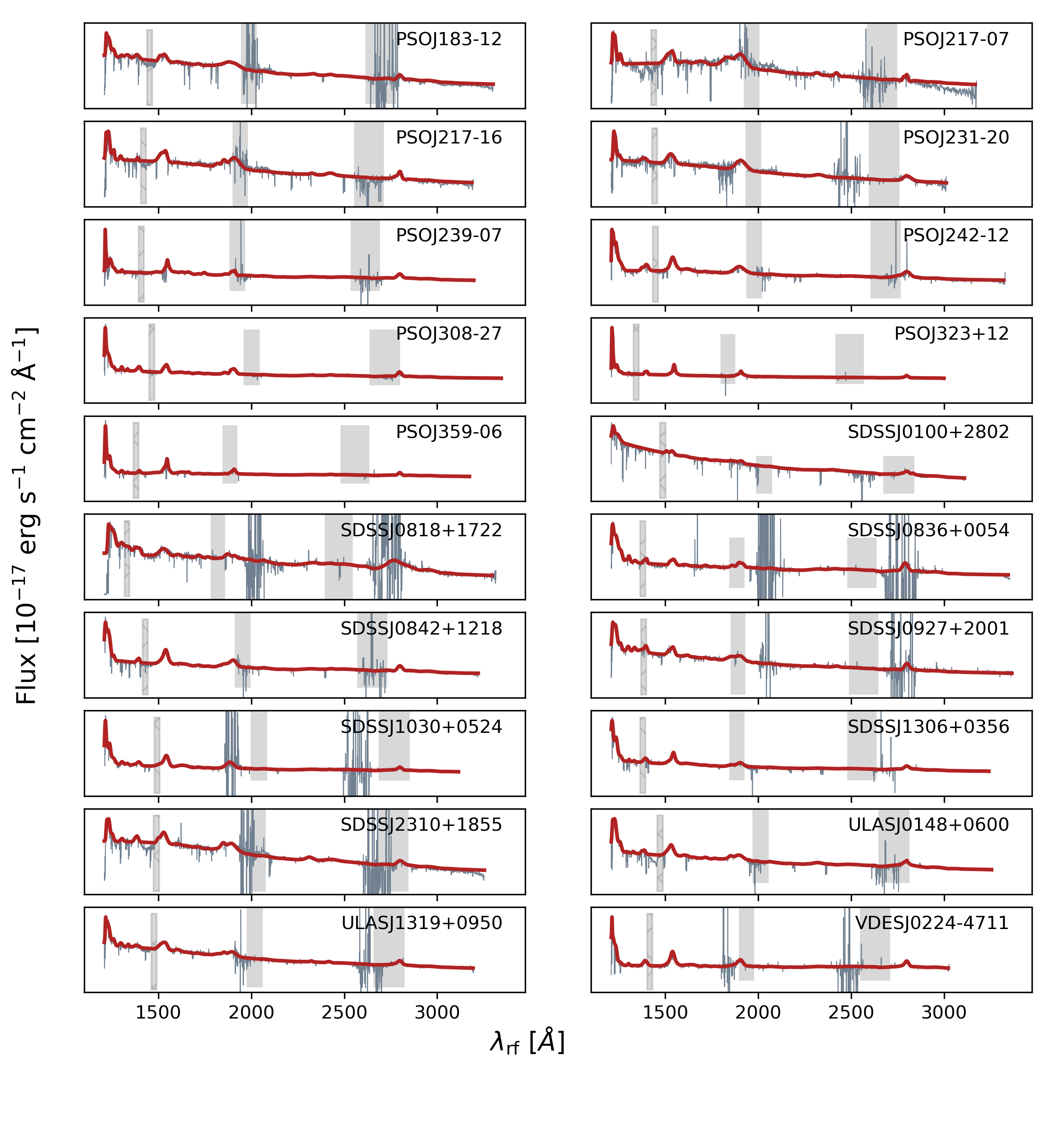}
    \caption{As in Figure \ref{fig:all-quasar-spectra1}.}
    \label{fig:all-quasar-spectra2}
\end{figure*}

In Figures \ref{fig:all-quasar-spectra1} and \ref{fig:all-quasar-spectra2}, we report the spectral fits to the continuum and emission lines (in red) for the entire E-XQR-30 sample (Section \ref{sec:spectral_modeling}). 
We note that BAL troughs are not modelled in these fits, and we refer to \cite{Bischetti_2022, Bischetti_2023} for a detailed analysis of BAL systems.
Gray-shaded regions, from left to right, highlight the overlap of X-Shooter VIS and NIR arms at observed wavelengths of 1000–1020 nm, and the spectral regions significantly affected by telluric absorption at 1350–1450 nm and 1800–1950 nm.

To assess whether truncating the \feii template bluewards of \civ affects the line decomposition, we select four sources from the EXQR-30 sample (two BAL quasars, J0439+1634 and ULASJ0148+0600, and two non-BAL quasars, ATLASJ029-36 and SDSSJ1306+0356)
%; Fig. \ref{fig:iron-comparison})
and repeat the spectral fitting using the original \feii template from \cite{Vestergaard+2001}, i.e. without the $\lambda < 1550$\AA\ cut.
To quantitatively compare the fits, we evaluate the $\chi^2$ over the $1300$–$1600$\AA\ interval, masking the BAL trough in BAL quasars, as they are not modelled. 
In all cases, the modified template is statistically preferred, yielding $\Delta\chi^2 = \chi^2_{\rm orig.} - \chi^2_{\rm mod.} > 30$ and $\Delta{\rm BIC} = {\rm BIC}_{\rm orig.} - {\rm BIC}_{\rm mod.} > 30$.
The resulting \civ FWHM values are consistent within $2\sigma$, indicating that the overall line profile is not significantly affected by the choice of \feii template. 
In contrast, this choice has an influence on \vel, as it modifies the blue end of the \civ line integration range. 
The effect is more pronounced in BAL quasars, where differences of $\sim 1000$~km s$^{-1}$ are found in \vel\ estimates, with the original template yielding smaller values. Nevertheless, since the truncated template is statistically favoured, we adopt the corresponding \vel\ measurements as our reference.

\begin{figure}
    \centering
    \includegraphics[width=0.6\linewidth]{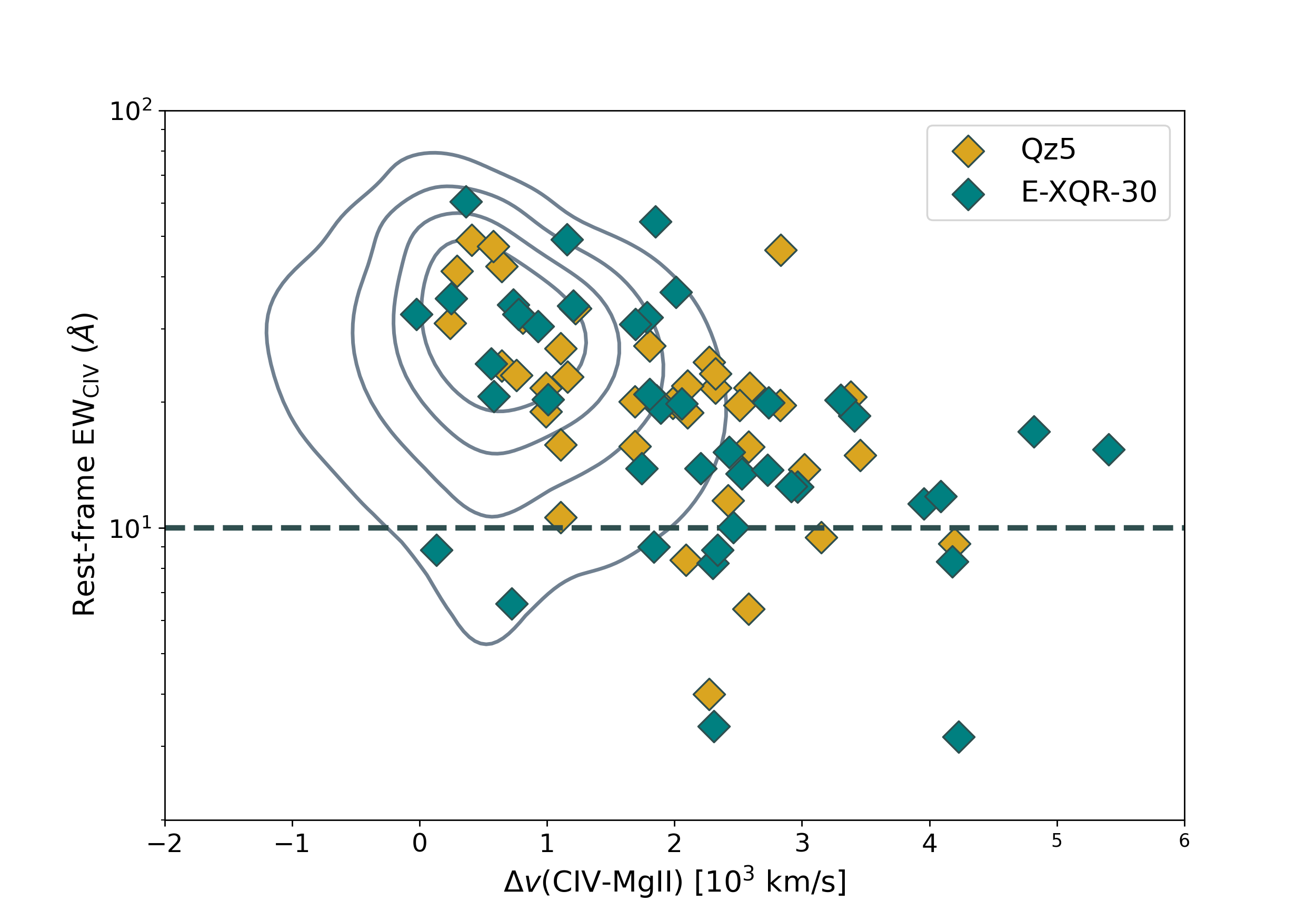}
    \caption{Rest-frame \civ EW as a function of \deltav\ for E-XQR-30 (teal) and Qz5 (golden) quasars. For the SDSS sample, we report the 0.15, 0.32, 0.50, and 0.68 density levels.
    The dark-gray dashed horizontal line marks EW$_{\rm CIV} = 10$~\AA, which is the threshold for the weak-line quasars regime.}
    \label{fig:ew-vs-velshift}
\end{figure}

In addition to \deltav(\civ-\mgii) and \vel, our spectral fitting allows us to derive quasar bolometric luminosities, the total \civ %and \mgii 
full widths at half maximum (FWHM), and rest-frame equivalent widths (EWs), all reported in Table \ref{tab:bal_highz_info}. 
Bolometric luminosities are derived from the monochromatic flux at $3000$~\AA\ using the bolometric correction of \cite{Richards+2006}: $\rm L_{bol} = 5.15 \lambda L_\lambda(3000\text{\AA})$~$\rm erg \ s^{-1}$. 
The reported uncertainties include only statistical errors ($\lesssim 0.1$~dex), but in the analysis we additionally include a systematic uncertainty of $0.3$~dex \citep[e.g., ][]{Richards+2006, Lai+2024, Brazzini+2025}. 
We note that the bolometric correction of \cite{Richards+2006} may overestimate $\rm L_{bol}$ by up to $\sim 0.13$~dex \citep[e.g.,][]{Runnoe+2012, Trakhtenbrot2012, Saccheo+2025}. Nevertheless, we retain this prescription to ensure consistency with previous studies, including \cite{Shen+2011} for SDSS. 
Our estimates of both $\rm L_{bol}$ and \civ FWHM are consistent within $3 \sigma$ with those of \cite{Mazzucchelli+2023}. 
A small number of \civ FWHM outliers show larger discrepancies, likely due to the presence of BAL troughs, which complicate the line fitting, and/or the proximity of \civ to the blue edge of the NIR arm, where the SNR is low.
The rest-frame EWs are estimated from best-fit line model integration. 
In Figure \ref{fig:ew-vs-velshift}, we report the \civ EW versus \deltav. These two quantities show a mild anti-correlation ($r=-0.140$, $p$-value~$=1.23 \times 10^{-7}$), in agreement with previous literature works \citep[e.g.,][]{Vietri+2018, Schindler+2020}. In particular, the high-redshift quasars occupy the same region in the diagram of the WISSH quasars analysed in \cite{Vietri+2018}.
The EW$_{\civ} < 10$~\AA\ regime denotes sources with significantly weaker (i.e small EW) \civ emission line than normal broad-line quasars \citep[weak-line quasars, WLQs; e.g.,][]{Wu+2011_WLQs, Luo+2015_WLQs}. 
Both EXQR-30 and Qz5 samples have some objects lying in this regime, for a total of 13 WLQs, of which five are BAL quasars (namely, PSOJ065+01, PSOJ089-15, J0923+0402, PSOJ183-12, and SDSSJ0836+0054). 

We further verify that the \civ FWHM and velocity shifts for the six BAL quasars from Qz5 that we re-analyse in this work are consistent with previous estimates from \cite{Brazzini+2025}, and find good agreement within $3 \sigma$. 
The redshifts and bolometric luminosities reported in Table \ref{tab:bal_highz_info} are from \cite{Brazzini+2025}. 

\begin{figure}
    \centering
    \includegraphics[width=0.7\linewidth]{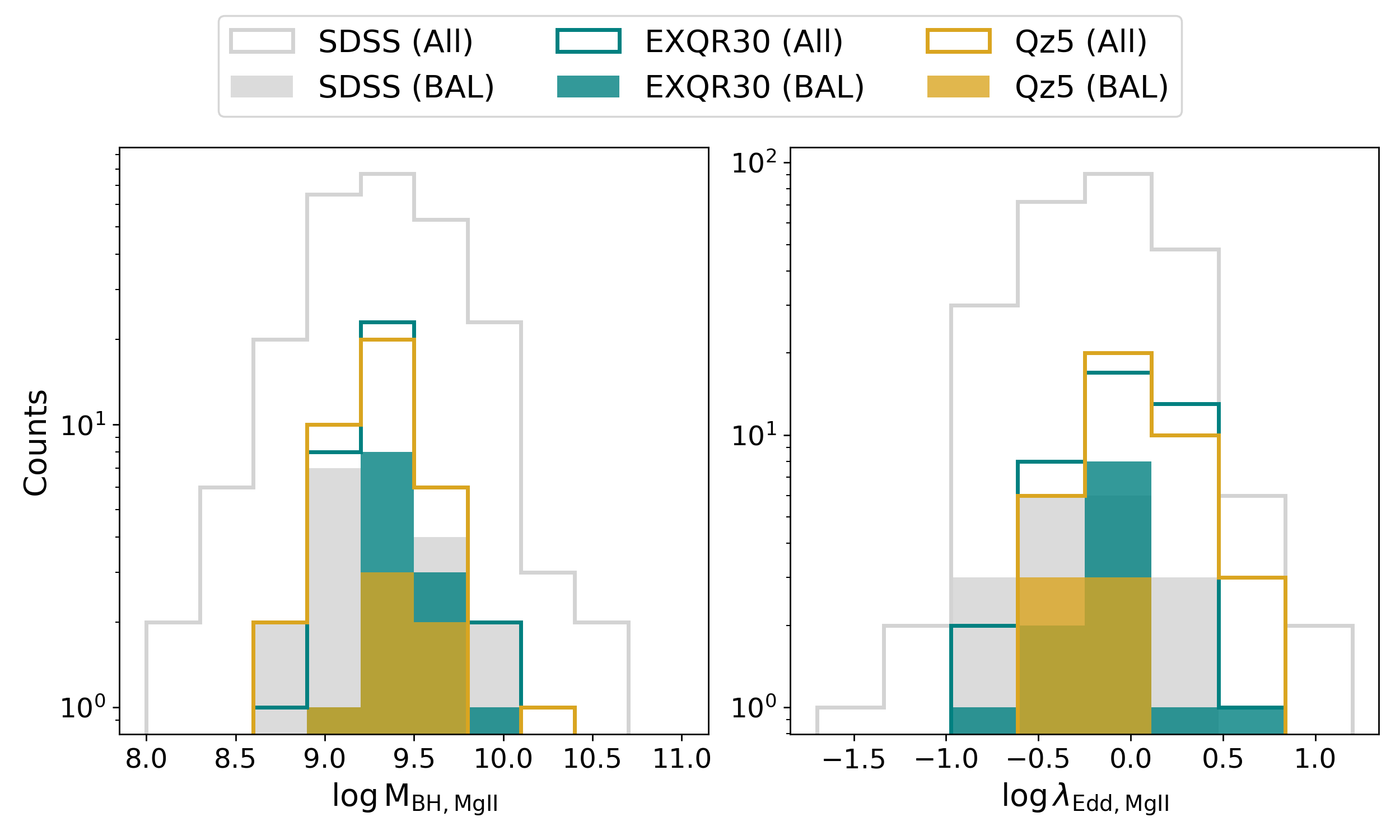}
    \caption{Black hole mass and Eddington ratio distributions from and \mgii diagnostics for E-XQR-30, Qz5, and SDSS datasets. The distributions of the entire samples satisfying the luminosity cut are displayed as empty histograms, while those of the BAL subsamples are displayed in filled histograms. The three distributions exhibit medians of -0.17, -0.01, and -0.02 for SDSS, Qz5 and E-XQR-30, respectively, while the median of the combined sample is -0.12.}
    \label{fig:BHmass_LEdd_hist}
\end{figure}

Lastly, we use monochromatic luminosities and line FWHM to estimate black hole masses, $\rm M_{BH}$, and Eddington ratios, \LEdd, using both \civ and \mgii estimators, following \cite{VestergaardPeterson2006_CIV} (with the correction by \citealp{Coatman+2017}) and \cite{VestergaardOsmer2009_MgII}, respectively. We report these results in Figure \ref{fig:BHmass_LEdd_hist}, where we integrate E-XQR-30 measurements with those for Qz5 and SDSS. 

The samples used all satisfy $  L_{\rm bol} > 10^{47}  $ erg/s, which preferentially selects objects with high $  L_{\rm bol}/M_{\rm BH}  $. 
Indeed, the peaks of the $  \log\lambda_{\rm Edd}  $ distributions 
%lie between $  -0.2  $ and $  -0.1  $, 
are of -0.17, -0.01 and -0.02 for the luminosity-cut SDSS, Qz5 and E-XQR-30 samples, respectively, whereas in the recent works of \cite{Risaliti+2026} and \cite{Fiore+2026} the peak of $  \log\lambda_{\rm Edd}  $ for luminosity-unbiased samples of type-I quasars is approximately $  -1  $. 
In those studies, the observed scatter in the $  \lambda_{\rm Edd}  $ distribution appears to be dominated by the typical uncertainty in single-epoch virial black-hole mass estimates ($  \sigma=0.3{-}0.4  $ dex), arising mainly from the unknown system inclination; the intrinsic scatter in the Eddington ratio is likely small. Objects viewed more face-on will have their black-hole masses underestimated (and $  \lambda_{\rm Edd}  $ correspondingly overestimated) when the virial estimator calibrated for an average inclination is applied, while the opposite occurs for more edge-on systems. Bright hyperluminous quasars tend to be viewed preferentially face-on, and therefore the true $  \lambda_{\rm Edd}  $ distributions are expected to be shifted toward lower values relative to those shown in Figure \ref{fig:BHmass_LEdd_hist}. 
Furthermore, BH masses of hyperluminous quasars can be underestimated also because radiati1on pressure can strongly modify the dynamics of BLR clouds \citep{Marconi+2008}. Nevertheless, the $  M_{\rm BH}  $ and $  \lambda_{\rm Edd}  $ distributions of BAL and non-BAL quasars are not significantly different for our samples.

\begin{table}
\centering
\small
\caption{Compilation of emission- and absorption-based outflow properties for the E-XQR-30 sample and for the six BAL quasars from Qz5. The table lists: redshift from \mgii peak, bolometric luminosity, \civ-\mgii velocity shift, total \civ FWHM and rest-frame EW, BAL properties (balnicity index, \vminBAL\ and \vmaxBAL) from \cite{Bischetti_2022, Bischetti_2023}, \vel\ and \velred\ (Sect. \ref{sec:analysis}). }
\renewcommand{\arraystretch}{1.1}
\begin{adjustbox}{width=\textwidth, max height=0.9\textheight}
\begin{tabular}{cccccccccccc}
\toprule
Quasar & $z$ & $\log L_{\mathrm{bol}}$ & $f$ & $\Delta v$(CIV--MgII) & FWHM$_{\mathrm{CIV}}$ & REW$_{\mathrm{CIV}}$ & BI & v$_{\mathrm{max,BAL}}$ & v$_{\mathrm{min,BAL}}$ & v$_{\mathrm{98}}$ & v$_{\mathrm{98,red}}$ \\
 &  & [erg/s] &  & [$10^3$~km/s] & [$10^3$~km/s] & [Å] & [km/s] & [$10^3$~km/s] & [$10^3$~km/s] & [$10^3$~km/s] & [$10^3$~km/s] \\
\midrule
ATLASJ025-33 & $6.334 \pm 0.015$ & $47.69 \pm 0.01$ & 1.31 & $2.74 \pm 1.02$ & $8.00^{+3.36}_{-1.29}$ & $20.0^{+5.5}_{-3.7}$ & - & - & - & $12.33 \pm 0.19$ & $8.11 \pm 0.01$ \\
ATLASJ029-36 & $5.997 \pm 0.003$ & $47.17^{+0.18}_{-0.23}$ & 1.29 & $0.56^{+0.19}_{-0.18}$ & $6.73^{+1.21}_{-1.01}$ & $24.8^{+2.6}_{-3.4}$ & - & - & - & $12.32 \pm 1.31$ & $7.57 \pm 0.22$ \\
CFHQSJ1509-1749 & $6.121 \pm 0.010$ & $47.42^{+0.03}_{-0.04}$ & 0.98 & $1.20^{+0.87}_{-0.58}$ & $5.56^{+2.32}_{-1.86}$ & $34.1^{+7.8}_{-8.5}$ & 300 & 28.84 & 23.27 & $6.92 \pm 0.06$ & $9.09 \pm 0.07$ \\
J0408-5632 & $6.028 \pm 0.004$ & $47.24^{+0.10}_{-0.09}$ & 1.21 & $2.20^{+0.26}_{-0.20}$ & $8.11^{+0.98}_{-0.68}$ & $13.9^{+1.0}_{-1.3}$ & 360 & 28.82 & 24.13 & $14.00 \pm 1.23$ & $8.44 \pm 0.11$ \\
J0439+1634$^{(a)}$ & $6.521 \pm 0.002$ & $48.34 \pm 0.01$ & 1.10 & $1.81^{+0.16}_{-0.15}$ & $5.70^{+0.33}_{-0.35}$ & $20.9^{+1.2}_{-1.6}$ & 2580 & 21.73 & 11.14 & $7.33 \pm 0.06$ & $6.17 \pm 0.03$ \\
J0923+0402 & $6.583 \pm 0.019$ & $47.50^{+0.05}_{-0.06}$ & 1.07 & $2.34^{+0.39}_{-1.84}$ & $4.08 \pm 0.41$ & $8.8^{+1.6}_{-1.1}$ & 13670 & 29.55 & 6.09 & $5.81 \pm 1.00$ & $5.72 \pm 0.02$ \\
J1535+1943 & $6.328 \pm 0.022$ & $47.56 \pm 0.06$ & 1.70 & $3.96^{+3.29}_{-1.69}$ & $10.55^{+2.24}_{-2.97}$ & $11.4^{+2.8}_{-3.1}$ & - & - & - & $14.09 \pm 0.31$ & $6.95 \pm 0.07$ \\
J2211-3206 & $6.333 \pm 0.004$ & $47.45 \pm 0.06$ & 1.13 & $1.74^{+0.50}_{-0.41}$ & $4.41^{+0.68}_{-0.84}$ & $13.9^{+3.5}_{-2.4}$ & 6840 & 24.34 & 9.24 & $6.43 \pm 0.26$ & $4.96 \pm 0.06$ \\
J2250-5015 & $5.971 \pm 0.003$ & $47.47 \pm 0.04$ & 1.16 & $5.41^{+0.45}_{-0.41}$ & $9.90^{+0.70}_{-0.97}$ & $15.4^{+1.8}_{-2.2}$ & 6330 & 50.32 & 22.50 & $12.88 \pm 0.48$ & $11.52 \pm 0.08$ \\
PSOJ007+04 & $5.991 \pm 0.004$ & $47.03^{+0.09}_{-0.10}$ & 0.97 & $1.89^{+0.39}_{-0.40}$ & $8.42^{+0.86}_{-0.72}$ & $19.4^{+1.6}_{-1.8}$ & - & - & - & $9.56 \pm 0.59$ & $9.37 \pm 0.07$ \\
PSOJ009-10 & $5.910 \pm 0.005$ & $47.36^{+0.09}_{-0.10}$ & 1.20 & $2.97^{+0.45}_{-0.42}$ & $9.98^{+0.83}_{-0.77}$ & $12.5^{+1.4}_{-1.7}$ & 1110 & 38.38 & 33.04 & $11.55 \pm 1.05$ & $8.61 \pm 0.20$ \\
PSOJ023-02 & $5.817 \pm 0.002$ & $47.34 \pm 0.04$ & 1.13 & $0.74^{+0.11}_{-0.09}$ & $4.93^{+0.13}_{-0.17}$ & $34.2^{+1.5}_{-1.3}$ & 2120 & 21.24 & 3.95 & $4.83 \pm 0.09$ & $4.95 \pm 0.05$ \\
PSOJ025-11 & $5.802 \pm 0.005$ & $47.31^{+0.05}_{-0.06}$ & 1.17 & $0.58^{+0.74}_{-0.55}$ & $5.77^{+2.08}_{-1.06}$ & $20.7^{+6.6}_{-4.0}$ & - & - & - & $7.88 \pm 0.67$ & $5.00 \pm 0.04$ \\
PSOJ029-29 & $5.979 \pm 0.002$ & $47.53 \pm 0.04$ & 1.12 & $2.06^{+0.49}_{-0.25}$ & $8.04^{+0.81}_{-0.94}$ & $19.8^{+1.9}_{-1.8}$ & - & - & - & $9.66 \pm 0.14$ & $7.59 \pm 0.05$ \\
PSOJ036+03 & $6.506 \pm 0.003$ & $47.44^{+0.05}_{-0.07}$ & 1.01 & $4.82^{+0.27}_{-0.20}$ & $9.86^{+0.28}_{-0.36}$ & $17.0 \pm 0.7$ & - & - & - & $12.11 \pm 0.16$ & $11.95 \pm 0.05$ \\
PSOJ060+24 & $6.169 \pm 0.002$ & $47.24^{+0.20}_{-0.38}$ & 1.07 & $0.78^{+0.14}_{-0.12}$ & $4.93 \pm 0.35$ & $32.5^{+4.5}_{-3.9}$ & - & - & - & $6.09 \pm 0.39$ & $5.11 \pm 0.12$ \\
PSOJ065+01 & $5.780 \pm 0.003$ & $47.16^{+0.13}_{-0.22}$ & 1.46 & $2.30^{+0.42}_{-0.41}$ & $5.89^{+2.73}_{-1.32}$ & $8.2^{+2.4}_{-1.6}$ & 920 & 38.61 & 16.89 & $8.13 \pm 2.01$ & $5.19 \pm 0.10$ \\
PSOJ065-26 & $6.179 \pm 0.005$ & $47.34^{+0.08}_{-0.09}$ & 1.57 & $4.23^{+0.88}_{-0.85}$ & $5.85^{+2.90}_{-2.09}$ & $3.2^{+1.5}_{-1.2}$ & - & - & - & $12.29 \pm 1.08$ & $7.06 \pm 0.10$ \\
PSOJ089-15 & $5.964 \pm 0.008$ & $47.57 \pm 0.06$ & 0.98 & $0.72^{+0.47}_{-0.36}$ & $2.13^{+0.49}_{-0.43}$ & $6.6^{+1.7}_{-1.4}$ & 13910 & 30.36 & 2.25 & $1.76 \pm 0.02$ & $2.52 \pm 0.06$ \\
PSOJ108+08 & $5.958 \pm 0.003$ & $47.51 \pm 0.04$ & 1.39 & $3.41^{+0.90}_{-0.89}$ & $9.51^{+1.42}_{-1.19}$ & $18.6^{+2.8}_{-2.9}$ & - & - & - & $11.42 \pm 0.18$ & $8.27 \pm 0.16$ \\
PSOJ158-14 & $6.058 \pm 0.004$ & $47.56 \pm 0.15$ & 1.14 & $2.01^{+0.24}_{-0.31}$ & $7.12^{+0.32}_{-0.29}$ & $36.7^{+1.7}_{-1.4}$ & - & - & - & $11.04 \pm 0.35$ & $7.96 \pm 0.15$ \\
PSOJ183+05 & $6.425 \pm 0.004$ & $47.13^{+0.12}_{-0.14}$ & 2.06 & $2.92^{+0.37}_{-0.33}$ & $8.92^{+1.02}_{-1.01}$ & $12.6 \pm 1.6$ & - & - & - & $12.48 \pm 0.17$ & $6.60 \pm 0.07$ \\
PSOJ183-12 & $5.889 \pm 0.005$ & $47.43^{+0.05}_{-0.06}$ & 1.29 & $4.18^{+0.25}_{-0.32}$ & $7.64^{+0.52}_{-0.48}$ & $8.3 \pm 0.7$ & 540 & 25.91 & 16.28 & $9.85 \pm 0.36$ & $8.15 \pm 0.16$ \\
PSOJ217-07 & $6.182 \pm 0.007$ & $47.19 \pm 0.08$ & 1.79 & $2.43^{+1.26}_{-1.08}$ & $8.39^{+3.38}_{-1.96}$ & $15.2^{+5.3}_{-3.7}$ & 4410 & 43.97 & 22.00 & $10.52 \pm 0.09$ & $5.64 \pm 0.05$ \\
PSOJ217-16 & $6.140 \pm 0.003$ & $47.27 \pm 0.06$ & 1.92 & $2.53^{+3.18}_{-0.55}$ & $8.83^{+1.26}_{-1.25}$ & $13.5^{+2.0}_{-2.5}$ & - & - & - & $10.78 \pm 0.08$ & $5.59 \pm 0.05$ \\
PSOJ231-20 & $6.553 \pm 0.004$ & $47.30^{+0.13}_{-0.23}$ & 1.12 & $4.45^{+0.19}_{-0.20}$ & $7.90^{+0.30}_{-0.40}$ & $11.5^{+0.8}_{-0.7}$ & 600 & 2.68 & 0.54 & $9.14 \pm 0.32$ & $10.13 \pm 0.11$ \\
PSOJ239-07 & $6.107 \pm 0.006$ & $47.49^{+0.04}_{-0.06}$ & 0.83 & $-0.03^{+0.27}_{-0.26}$ & $3.78^{+0.65}_{-0.49}$ & $32.5^{+4.9}_{-3.3}$ & - & - & - & $3.99 \pm 0.04$ & $5.02 \pm 0.13$ \\
PSOJ242-12 & $5.854 \pm 0.005$ & $47.25^{+0.07}_{-0.08}$ & 0.95 & $1.69^{+0.26}_{-0.31}$ & $6.24^{+0.97}_{-0.86}$ & $30.8^{+6.9}_{-7.9}$ & - & - & - & $6.61 \pm 0.16$ & $7.56 \pm 0.13$ \\
PSOJ308-27 & $5.798 \pm 0.002$ & $47.39 \pm 0.03$ & 1.21 & $1.01^{+0.26}_{-0.29}$ & $5.51^{+0.68}_{-0.66}$ & $20.3^{+2.4}_{-2.0}$ & - & - & - & $6.51 \pm 0.27$ & $4.96 \pm 0.04$ \\
PSOJ323+12 & $6.584 \pm 0.002$ & $47.26^{+0.09}_{-0.12}$ & 1.06 & $0.25^{+0.07}_{-0.09}$ & $2.82^{+0.09}_{-0.11}$ & $35.5^{+1.6}_{-2.1}$ & - & - & - & $5.14 \pm 0.09$ & $5.42 \pm 0.03$ \\
PSOJ359-06 & $6.168 \pm 0.003$ & $47.31^{+0.05}_{-0.07}$ & 1.30 & $0.36^{+0.12}_{-0.14}$ & $3.25^{+0.38}_{-0.23}$ & $60.5^{+4.2}_{-3.4}$ & - & - & - & $7.39 \pm 0.05$ & $5.20 \pm 0.06$ \\
SDSSJ0100+2802 & $6.316 \pm 0.005$ & $48.16 \pm 0.01$ & 1.52 & $2.31^{+0.38}_{-0.26}$ & $9.36 \pm 0.56$ & $3.3 \pm 0.4$ & - & - & - & $10.00 \pm 0.03$ & $5.61 \pm 0.02$ \\
SDSSJ0818+1722 & $5.858 \pm 0.020$ & $47.61^{+0.02}_{-0.03}$ & 0.99 & $2.46^{+0.65}_{-0.99}$ & $9.81^{+0.72}_{-0.83}$ & $10.0 \pm 0.9$ & - & - & - & $8.95 \pm 0.25$ & $10.58 \pm 0.40$ \\
SDSSJ0836+0054 & $5.780 \pm 0.012$ & $47.85 \pm 0.01$ & 1.34 & $0.13^{+0.90}_{-0.53}$ & $6.51^{+0.68}_{-1.11}$ & $8.9 \pm 1.5$ & 160 & 28.04 & 25.48 & $7.28 \pm 0.24$ & $5.13 \pm 0.08$ \\
SDSSJ0842+1218 & $6.062 \pm 0.001$ & $47.25^{+0.06}_{-0.07}$ & 1.15 & $1.78 \pm 0.11$ & $6.07^{+0.41}_{-0.45}$ & $32.0^{+3.2}_{-3.7}$ & 740 & 27.38 & 19.92 & $9.89 \pm 0.14$ & $7.86 \pm 0.10$ \\
SDSSJ0927+2001 & $5.763 \pm 0.003$ & $47.14^{+0.06}_{-0.04}$ & 1.57 & $1.84^{+0.85}_{-0.71}$ & $5.69^{+2.00}_{-1.51}$ & $9.0^{+2.9}_{-2.7}$ & - & - & - & $8.66 \pm 0.28$ & $4.95 \pm 0.05$ \\
SDSSJ1030+0524 & $6.306 \pm 0.002$ & $47.36^{+0.03}_{-0.05}$ & 1.24 & $1.16^{+0.22}_{-0.26}$ & $5.49^{+1.90}_{-0.95}$ & $49.1^{+15.4}_{-12.5}$ & - & - & - & $11.20 \pm 0.44$ & $6.29 \pm 0.03$ \\
SDSSJ1306+0356 & $6.019 \pm 0.002$ & $47.27^{+0.02}_{-0.03}$ & 1.10 & $0.93 \pm 0.08$ & $4.41 \pm 0.19$ & $30.4^{+1.4}_{-1.2}$ & - & - & - & $10.00 \pm 0.12$ & $8.98 \pm 0.09$ \\
SDSSJ2310+1855 & $5.993 \pm 0.003$ & $47.45^{+0.06}_{-0.10}$ & 1.20 & $4.09^{+0.35}_{-0.39}$ & $8.26^{+1.20}_{-1.76}$ & $11.9 \pm 1.1$ & 590 & 26.82 & 18.52 & $11.63 \pm 0.33$ & $9.21 \pm 0.13$ \\
ULASJ0148+0600 & $5.978 \pm 0.002$ & $47.47 \pm 0.01$ & 1.15 & $3.30^{+0.17}_{-0.20}$ & $7.06^{+0.50}_{-0.54}$ & $20.2^{+1.8}_{-2.2}$ & 2250 & 27.93 & 16.26 & $9.65 \pm 0.08$ & $8.61 \pm 0.04$ \\
ULASJ1319+0950 & $6.126 \pm 0.003$ & $47.33^{+0.02}_{-0.04}$ & 1.37 & $2.73^{+2.25}_{-0.92}$ & $8.34^{+2.83}_{-2.34}$ & $13.8^{+3.5}_{-4.6}$ & - & - & - & $11.49 \pm 0.06$ & $7.71 \pm 0.03$ \\
VDESJ0224-4711 & $6.527 \pm 0.002$ & $47.51^{+0.04}_{-0.05}$ & 1.04 & $1.85^{+0.06}_{-0.08}$ & $5.43^{+0.30}_{-0.24}$ & $54.2^{+5.5}_{-4.4}$ & - & - & - & $9.56 \pm 0.39$ & $9.62 \pm 0.03$ \\
\midrule
J2225+0330 & $5.2523 \pm 0.0004$ & $47.20 \pm 0.18$ & 1.50 & $0.99^{+0.72}_{-0.57}$ & $4768^{+1980}_{-1281}$ & $18.2^{+6.5}_{-4.8}$ & 5390 & 23.58 & 9.46 & $7.87 \pm 0.42$ & $3.51 \pm 0.16$ \\
J0957+1016 & $5.1274 \pm 0.0011$ & $47.08 \pm 0.14$ & 1.18 & $2.24^{+0.44}_{-0.65}$ & $9987^{+1614}_{-2099}$ & $33.5^{+5.0}_{-4.6}$ & 5680 & 23.12 & 9.12 & $8.60 \pm 0.07$ & $7.05 \pm 0.64$ \\
J1004+2025 & $5.0634 \pm 0.0014$ & $47.24 \pm 0.24$ & 1.55 & $4.28^{+8.27}_{-3.50}$ & $14191^{+7387}_{-5995}$ & $41.1^{+17.2}_{-18.6}$ & 6080 & 48.67 & 24.87 & $19.32 \pm 1.65$ & $8.43 \pm 0.64$ \\
J1601-1828 & $5.0502 \pm 0.0006$ & $47.41 \pm 0.04$ & 1.58 & $3.22^{+0.72}_{-0.54}$ & $8247^{+1132}_{-1459}$ & $24.4^{+3.3}_{-3.1}$ & 2610 & 40.98 & 15.22 & $9.39 \pm 0.08$ & $5.42 \pm 0.18$ \\
J0338+0021 & $5.0230 \pm 0.0005$ & $47.15 \pm 0.22$ & 2.10 & $1.21^{+0.44}_{-0.22}$ & $7367^{+578}_{-457}$ & $11.8^{+1.3}_{-0.9}$ & 890 & 22.49 & 12.41 & $9.49 \pm 0.41$ & $3.30 \pm 0.20$ \\
J0017-1000 & $5.0170 \pm 0.0020$ & $47.35 \pm 0.10$ & 1.14 & $2.67^{+0.11}_{-0.10}$ & $7074^{+183}_{-126}$ & $28.5^{+0.8}_{-1.2}$ & 4430 & 21.14 & 9.57 & $8.86 \pm 0.06$ & $6.25 \pm 0.11$ \\
\bottomrule
\end{tabular}
\end{adjustbox}
\parbox{\textwidth}{\vspace{2mm}
\textbf{Notes:} 
\textit{(a)} This quasar is gravitationally lensed \citep{Fan+2019}; the reported bolometric luminosity is not corrected for magnification.}
\label{tab:bal_highz_info}
\end{table}

\section{BAL quasars in E-XQR-30 and Qz5}

\begin{figure*}
    \centering
    \includegraphics[width = \linewidth]{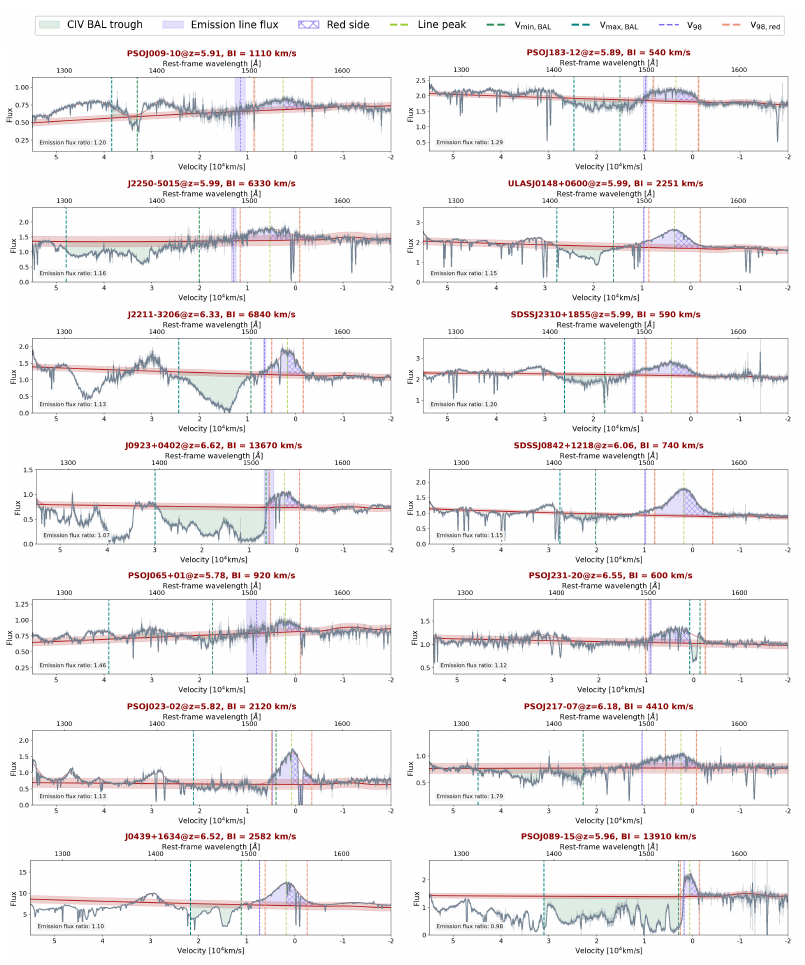}
    \caption{Spectral regions encompassing \civ BAL trough and \civ emission in E-XQR-30 quasars with $\rm BI > 500$~km s$^{-1}$. The red solid curve represents the continuum (Sect. \ref{sect:continuum_emission_modelling}), and the red-shaded area is the $\pm 1 \sigma$ spectral uncertainty used for line profile integration (light purple; Sect. \ref{sec:analysis}) and \vel\ (purple dashed line) measurement. The red side over which \velred\ (pink dotted line) is estimated is highlighted with hatches. PSOJ089-15 is the only BAL quasar in E-XQR-30 with $f<1$, for which we use \velred. 
    %In the others, \velred\ is shown for comparison but not used in our analysis. 
    PSOJ231-20 is a peculiar source exhibiting a blue-shifted \civ emission line (by $\sim 4440$~km s$^{-1}$) and a narrow BAL trough showing no motion with respect to the systemic redshift. \vminBAL\ and \vmaxBAL \citep{Bischetti_2022, Bischetti_2023} are also reported for reference. 
    }
  \label{fig:EXQR30-BALs}
\end{figure*}

\begin{figure*}
    \centering
    \includegraphics[width = \linewidth]{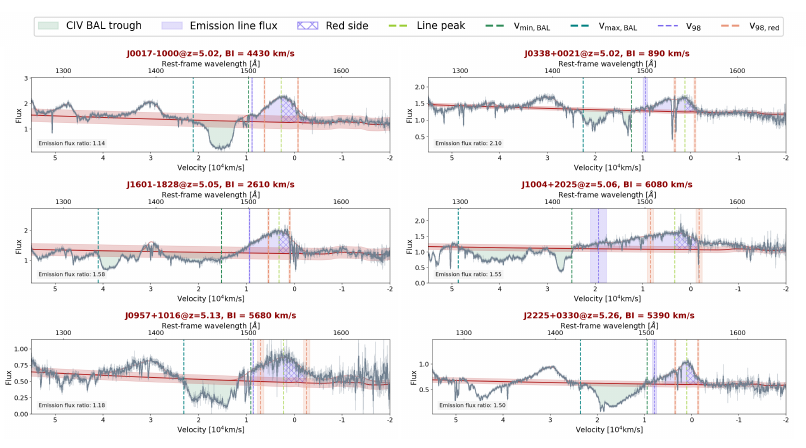}
    \caption{Same as Fig. \ref{fig:EXQR30-BALs}, but for the Qz5 sample. \vminBAL\ and \vmaxBAL\ are taken from \cite{Brazzini+2025}. }
  \label{fig:Qz5-BALs}
\end{figure*}

We report in Figures \ref{fig:EXQR30-BALs} and \ref{fig:Qz5-BALs} a zoom-in of the \civ spectral region for the BAL quasars in the E-XQR-30 and Qz5 samples with $\rm BI > 500$~km s$^{-1}$.
Under the sign convention in the figures, negative velocities are associated to outflows, and the larger the absolute value, the faster the wind. 
We report in dark gray the quasar spectra, shifted to rest-frame assuming the \mgii-peak redshift as systemic. The light-green dashed vertical line highlights the peak position of \civ line, therefore representing \deltav(\civ-\mgii). 
The red solid line is the continuum bestfit model, including the power-law, the polynomial correction, the Balmer pseudo-continuum, and the \feii template. 
The red-shaded region represents the $\pm 1 \sigma$ noise on the continuum, adopted as the threshold for the emission line integration (see Sect. \ref{sect:continuum_emission_modelling} for more details). 
In light purple we report the \civ-line integration region, with the red side highlighted with hatches. 
The \vel\ estimate is reported as a light purple vertical dashed line, and \velred\ is reported in pink. 
We also display the BAL velocities in green, and highlight in light green the BAL trough delimited by these velocities, but we also remind that these estimates are taken from \cite{Bischetti_2022, Bischetti_2023}, calculated over a different spectral template from the red one reported here. 
Furthermore, in cases where our \mgii-based redshift estimates differ significantly from those reported in the literature by \cite{Mazzucchelli+2023}, the BAL velocities must be corrected accordingly, as the change in redshift implies a shift in the adopted zero-velocity reference.

\section{Relevant sources in the SDSS sample}

\begin{figure*}
    \centering
    \includegraphics[width=\linewidth,trim=0 3cm 0 0, clip]{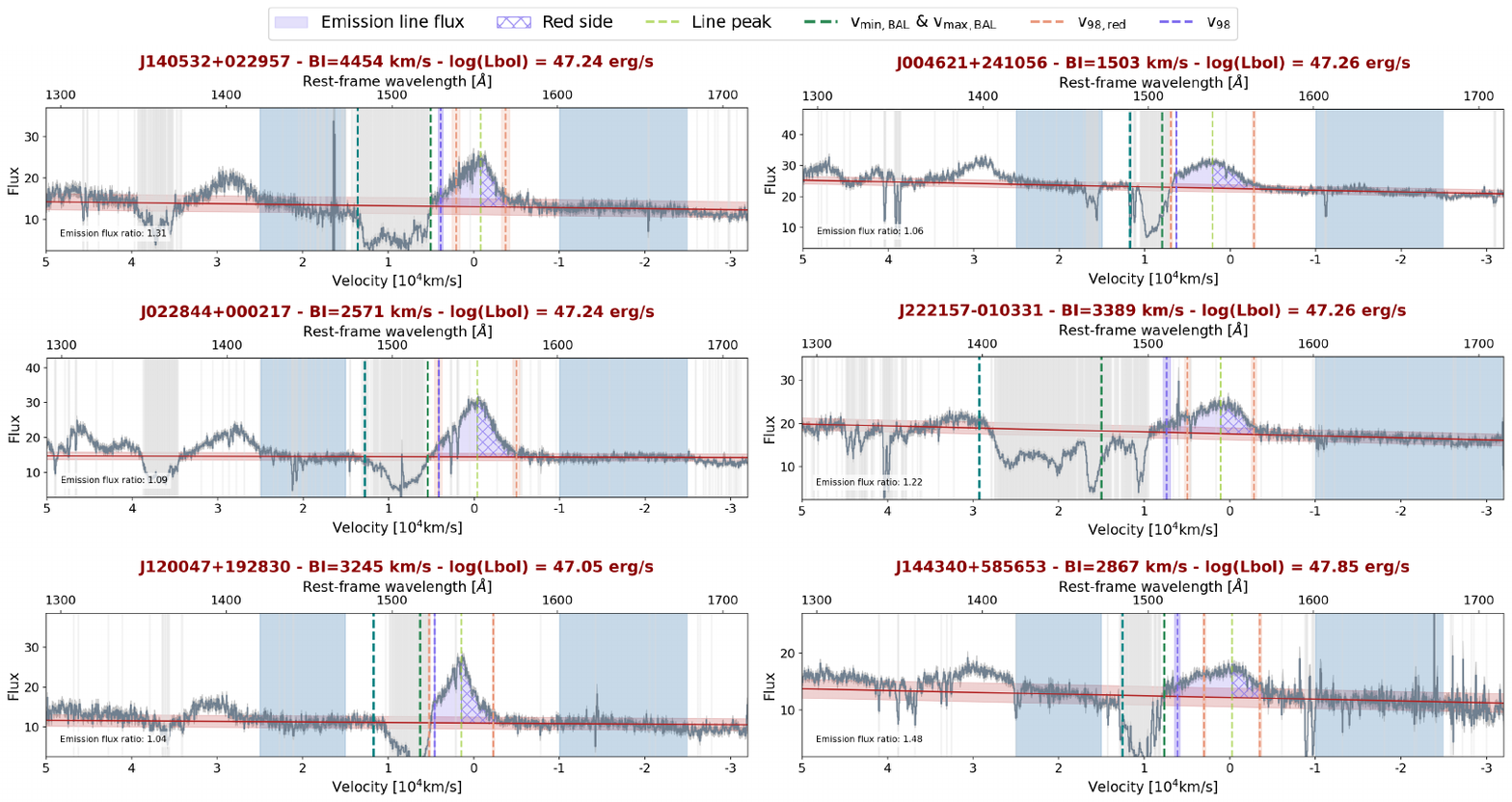}
    \caption{Some examples of BAL quasars from SDSS with $\rm \log L_{bol}>10^{47}$ erg s$^{-1}$ and $f>1$.
    The regions for continuum fitting are highlighted in light blue, and the resulting continuum estimate is reported in red, with the $\pm 1 \sigma$ uncertainty reported in light red.
    The \civ line used to calculate \vel\ is shown as a light-purple shaded region, with the corresponding \vel\ indicated by the dashed purple vertical line. The hatched region marks the red side of the line used to estimate \velred, which is indicated by the dashed salmon vertical line.
    \vminBAL\ and \vmaxBAL\ are reported as dashed green vertical lines. Spectral regions affected by absorption, excluded from the integration procedure, are highlighted in gray.}
  \label{fig:sdss-BALs-f_higher_1}
\end{figure*}

\begin{figure*}
    \centering
    \includegraphics[width=\linewidth]{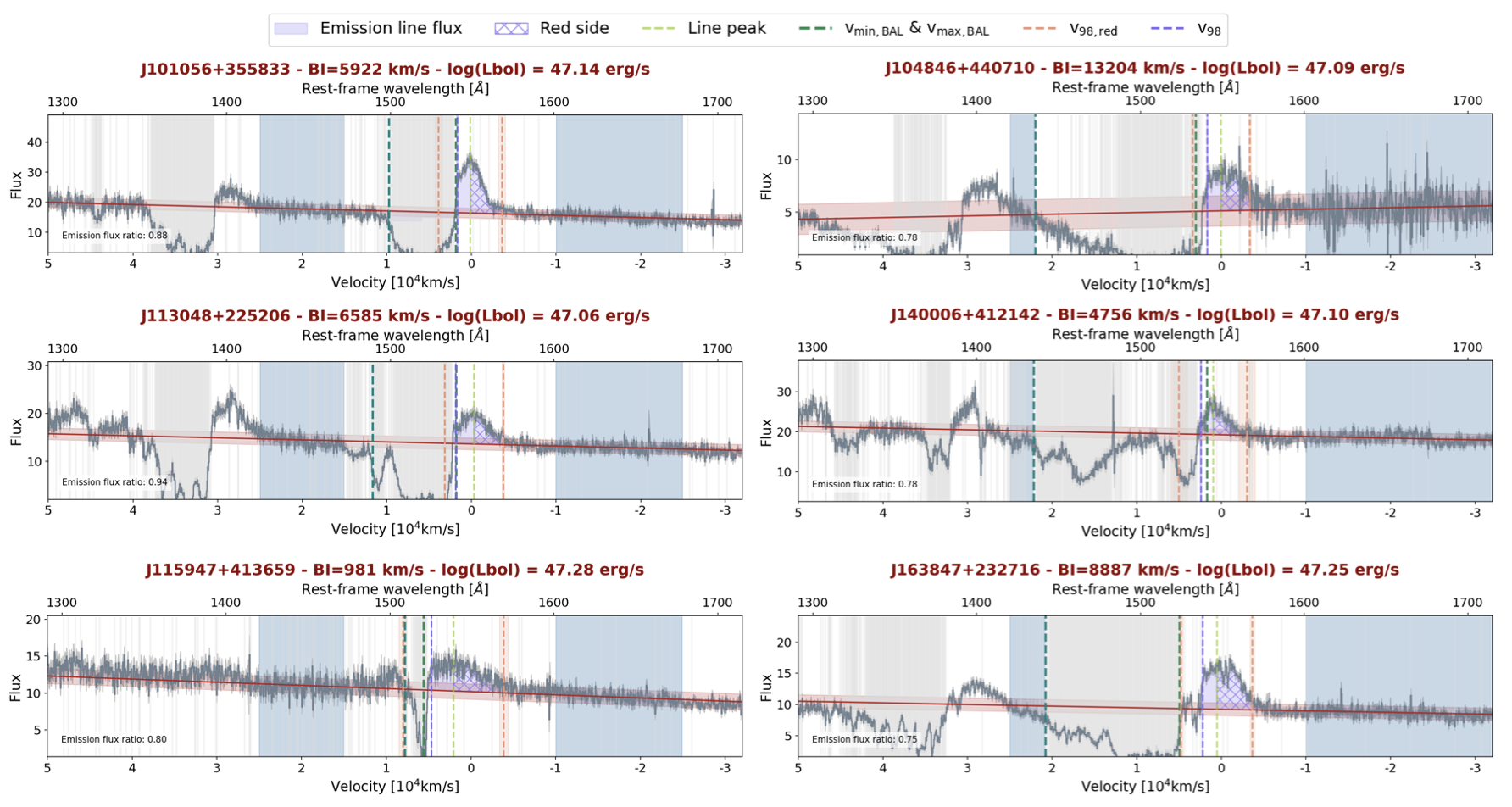}
    \caption{Same of Fig. \ref{fig:sdss-BALs-f_higher_1}, but for SDSS BAL quasars with $f<1$.}
  \label{fig:sdss-BALs-f_lower_1}
\end{figure*}

\begin{figure*}
    \centering
    \includegraphics[width=\linewidth]{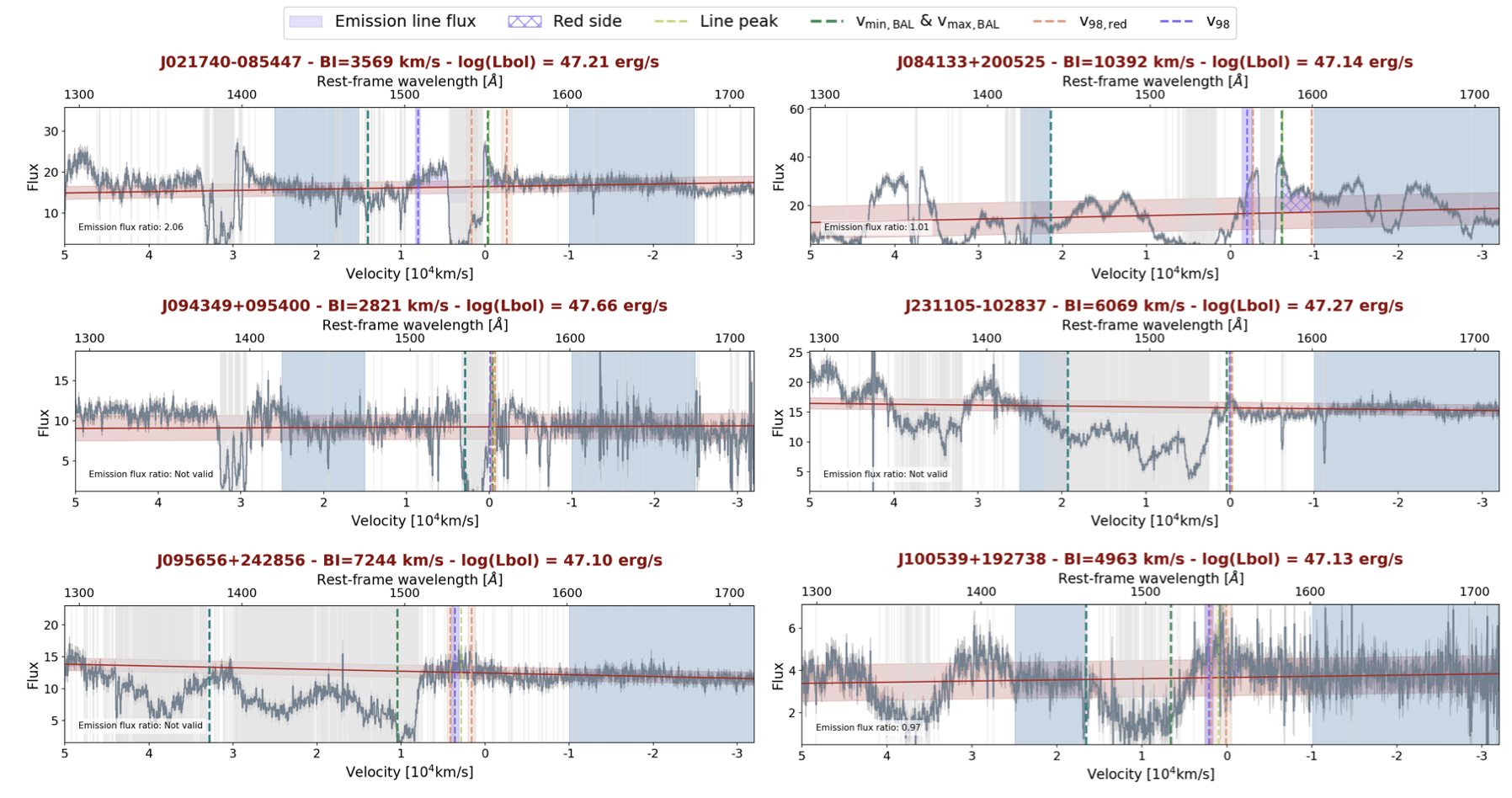}
    \caption{Same of Figs. \ref{fig:sdss-BALs-f_higher_1}, \ref{fig:sdss-BALs-f_lower_1}, but for SDSS BAL quasars excluded from our analysis via visual inspection.}
  \label{fig:sdss-BALs-excluded}
\end{figure*}

We display some relevant examples from the SDSS sample, including BAL quasars with either clear \civ emission line profiles ($f > 1$, Fig. \ref{fig:sdss-BALs-f_higher_1}), or with the profile significantly absorbed by the BAL trough ($f < 1$, Fig. \ref{fig:sdss-BALs-f_lower_1}), for which we adopt \vel\ and \velred, respectively, as emission-based outflow velocity estimates. 
In Figure \ref{fig:sdss-BALs-excluded}, instead, we report some critical sources excluded from the analysis because of the extreme contamination of \civ emission by the BAL trough, which makes the identification of the line peak and the subsequent \vel\ calculation unreliable. 
The minimum and maximum BAL velocities are taken from \cite{Bischetti_2022, Bischetti_2023}. 

\section{Analytic toy model for clumpy outflows}
\label{Appendix:toymodel}

\begin{figure}
    \centering
    \includegraphics[width=0.7\linewidth]{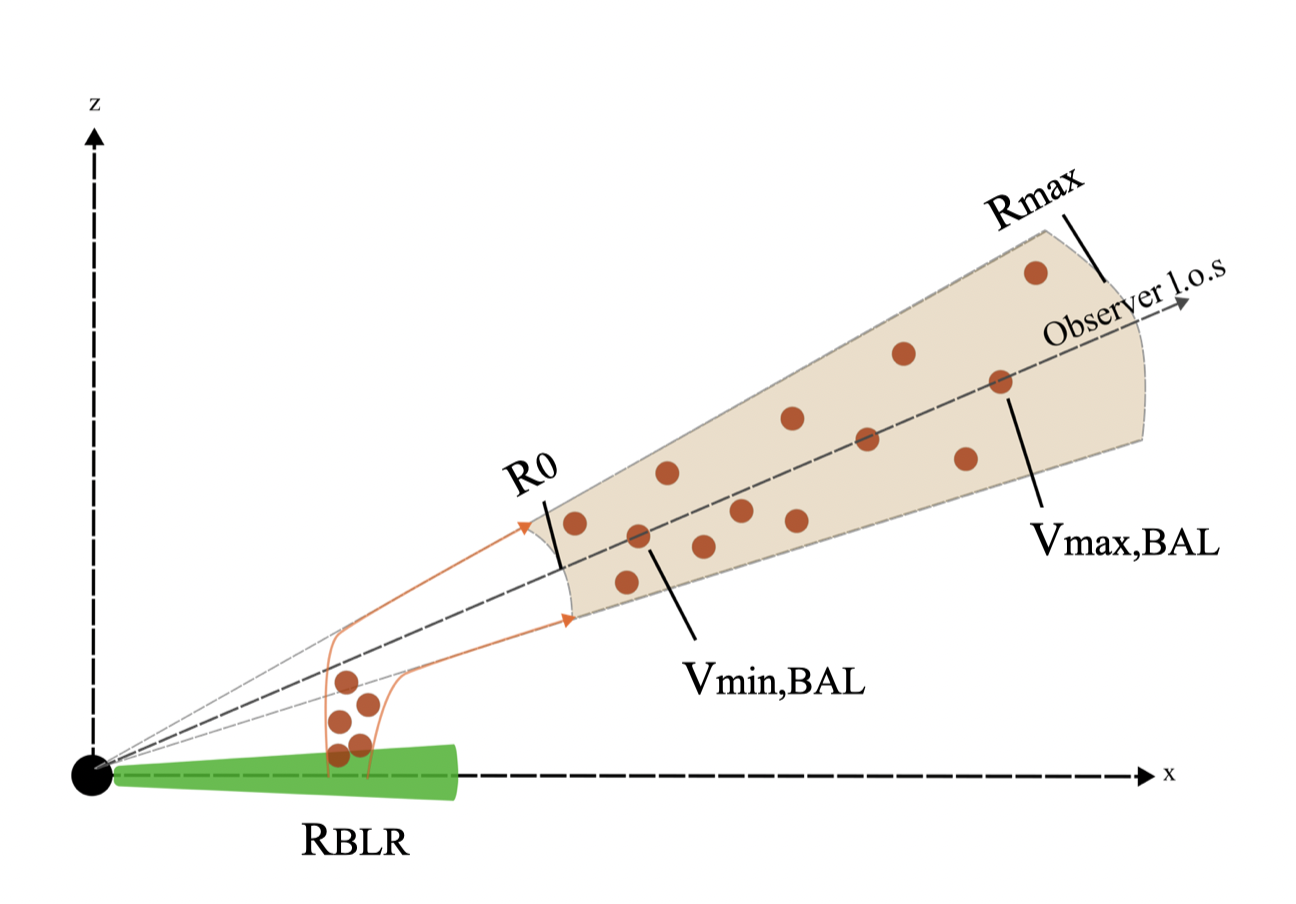}
    \caption{Cartoon illustrating the geometry adopted for the clumpy outflow toy model. The wind is launched from the BLR at $\rm R_{\rm{BLR}}$, initially rising vertically above the accretion disc before being accelerated radially outward by radiation pressure. We assume that at $\rm R_0$ the clumps reach an ionisation state sufficiently low to become optically thick to \civ absorption, thereby defining the inner radius of the BAL outflow. This radius also corresponds to the location where the wind velocity is set to $\rm v_0$ in the model. The wind is modelled as a population of discrete clumps whose number density follows the underlying wind density profile. The subset of clumps intersecting the observer’s line of sight determines the observed values of \vminBAL\ and \vmaxBAL.}
    \label{fig:toymodel_geometry}
\end{figure}

As discussed in Section \ref{sec:Results:clumpyoutflows}, we interpret the observed \vminBAL--\vmaxBAL\ relation within a clumpy outflow scenario, inspired by the classical quasar wind model of \cite{Elvis2000}.
We assume a bi-conical BAL outflow composed of discrete, identical clumps with mass $\rm M_c$, originating from the accretion disk. 
These clumps initially rise vertically above the disk, before being accelerated radially outward by radiation pressure (Fig.\,\ref{fig:toymodel_geometry}). 
We further assume that the observed absorption traces only the radially accelerated portion of the wind, while the dense material near the disk corresponds instead to BLR-scale winds.
This geometry is consistent with the scenario proposed by \cite{Vietri+2018}, who found evidence for a polar configuration of BLR gas clouds.
In reality, BAL winds are likely highly inhomogeneous media consisting of both `smooth' diffuse gas and clumps, as suggested by the wide diversity of absorption profiles observed in Figs.\,\ref{fig:EXQR30-BALs}, \ref{fig:Qz5-BALs}, \ref{fig:sdss-BALs-f_higher_1} and \ref{fig:sdss-BALs-f_lower_1}. 
However, modelling such a complexity is beyond the scope of this simplified treatment, and in what follows we assume that all of the wind mass is contained within the clumps. 

We consider a steady-state wind continuously accelerated by radiation pressure, producing a radial velocity profile that increases with distance from the black hole.
We parametrise the velocity field in terms of the terminal velocity, $\rm v_{\infty}$, following the standard prescription commonly adopted in radiation-driven wind models \citep[e.g.,][]{Knigge+1995, Sim+2008, Hagino+2015, Hagino+2016, Matzeu+2022}: 
\begin{equation}
    \label{eqn:v_l}
   \rm  v_l(R) = v_0 + (v_{\infty} - v_0) \left(1 - \frac{R_v}{R_v + (R-R_0)} \right)^{\beta}.
\end{equation}

Here, $\rm R$ is the radial distance from the black hole, $\rm R_0$ is the launch radius of the wind, $\rm v_0$ is the wind velocity at $\rm R_0$, $\rm R_v$ is the velocity scale length at which the wind reaches $\rm 0.5 v_{\infty}$, and $\beta$ is the velocity exponent, that we fix at 1.
%\citep{Hagen+2026}. 
Radiation-driven winds are expected to accelerate rapidly, thus we adopt $\rm R_v = R_0$, implying that the outflow reaches $\rm 0.5v_{\infty}$ at $\rm 2R_0$.

We assume inner and outer radial boundaries of $\rm R_0 = 10^4\,R_{\rm G}$ and $\rm R_{\rm max} = 5\cdot 10^7\,R_{\rm G}$, where $\rm R_{\rm G}=GM/c^2$. 
For a black hole mass of $\rm 10^{9.4}\,M_\odot$, corresponding to the average value of the E-XQR-30 sample, these limits translate into a radial range of $\sim 1$--$6000$\,pc, consistent with the typical spatial scales inferred for BAL outflows \citep[e.g.][]{Arav+2012, Arav+2013, Arav+2018, Arav+2020_specific_single_source, Hemler+2019, Miller+2020, He+2022, Bischetti+2024}. 
We further adopt inner and terminal velocities of $\rm v_0 = 10^3\,\rm km\,s^{-1}$ and $\rm v_{\infty} = 6\times10^4\,\rm km\,s^{-1} \simeq 0.2c$, respectively. The lower limit reflects the fact that, by construction (see Section~\ref{sec:analysis}), our sample does not include sources with $\vminBAL \leq \vel$, while the upper limit is consistent with the maximum velocities expected for radiation-driven winds once relativistic effects are taken into account \citep{Luminari+2021}.

To estimate the \vminBAL\ and \vmaxBAL\ distributions, we first derive the radial probability density function of the clumps, $\rm p_c(R)$, from the wind density profile \ref{eq:rho}. 
We then randomly draw $\rm N_c$ clumps intersecting the observer’s line of sight, and quantify \vminBAL\ and \vmaxBAL\ as the velocities associated to the clumps closest to and furthest from the central black hole, respectively.
Repeating this procedure $\rm N_{\rm obs}$ times provides an estimate of the probability distribution expected for the observed velocity samples.

We subdivide the wind into concentric radial shells with volume element $\rm dV=\Omega_w R^2dR$, where $\rm \Omega_w$ is the solid angle of the wind as seen from the central black hole.
The wind mass within each shell is therefore $\rm dM_w = \rho(R)dV=\rho(R) \Omega_w R^2 dR$, where $\rm \rho(R)$ is the wind mass density. 
Assuming a steady-state outflow, the wind mass outflow rate $\rm dM_w/dt = \dot{M}_w$ is constant throughout the entire wind.
Using the wind velocity law from Eq. \ref{eqn:v_l}, we can parametrise it as $\rm \dot{M}_w=\rho(R)\Omega_wR^2v_l(R)$, and derive $\rm \rho(R)$: 
\begin{equation}
\label{eq:rho}
    \rm \rho(R) = \frac{\dot{M}_w}{\Omega_wR^2v_{l}(R)}.
\end{equation}

This translates into a clump number density profile $\rm n_c(R)= \rho(R)/M_c$, from which we can derive the clump radial probability density profile:

\begin{equation}
    \label{eqn:p_c}
    \rm p_c(R) = \frac{n_c(R)}{\int\limits_{R_0}^{\infty} n_c(R) dR} = \frac{1}{R^2v_l(R) \int\limits_{R_0}^{\infty} \frac{1}{R'^2v_l(R')} dR'},
\end{equation}

corresponding to the probability of finding a clump at radius $\rm R$.

We now estimate the \vminBAL\ and \vmaxBAL\ distributions by repeatedly drawing $\rm N_c$ clumps along the observer’s line of sight according to Eq. \ref{eqn:p_c}.
We leave $\rm N_c$ as an input parameter and tune it to qualitatively reproduce the observed velocity distributions, although in principle it could be estimated analytically by integrating $\rm n_c(R)$ over a cylindrical volume aligned with the line of sight for a given $\rm \dot{M}_w$ and clump size.
The best qualitative match is achieved for $\rm N_c=2$, and the resulting distribution is shown in Fig.\,\ref{fig:toymodel_vel_dist}. 
A single clump ($\rm N_c=1$) naturally produces a one-to-one relation, while increasing the number of clumps systematically shifts the distribution towards more extreme values of both \vminBAL\ and \vmaxBAL.

\end{appendix}
\end{document}